\documentclass[11pt]{article}

\usepackage[preprint]{acl}

\usepackage{times}
\usepackage{latexsym}
\usepackage[T1]{fontenc}
\usepackage[utf8]{inputenc}
\usepackage{microtype}
\usepackage{inconsolata}
\usepackage{graphicx}

\usepackage{booktabs}
\usepackage{multirow}
\usepackage{array}
\usepackage{amsmath,amssymb}
\usepackage{enumitem}
\usepackage{xcolor}
\usepackage{placeins}
\usepackage{tikz}
\usetikzlibrary{arrows.meta,positioning,calc}
\definecolor{PBlue}{HTML}{164DCF}
\definecolor{PBlueFill}{HTML}{EEF5FF}
\definecolor{POrange}{HTML}{EC790C}
\definecolor{POrangeFill}{HTML}{FFF3E8}
\definecolor{PGreen}{HTML}{337A35}
\definecolor{PGreenFill}{HTML}{F0F7EC}
\definecolor{PGray}{HTML}{56616D}
\definecolor{PGrayFill}{HTML}{F7F8FA}
\definecolor{Ink}{HTML}{17212B}
\tikzset{
 flow/.style={-{Stealth[length=2.6mm]},draw=black,line width=1.0pt,rounded corners=1.4mm},
 feedback/.style={flow,draw=POrange,line width=1.0pt},
 evalflow/.style={flow,dashed,draw=PGreen},
 band/.style={rounded corners=1.6mm,minimum height=.42cm,inner sep=1.5mm,align=center,font=\sffamily\fontsize{10}{11}\selectfont},
 title/.style={align=center,font=\sffamily\bfseries\fontsize{10.5}{11.8}\selectfont,execute at begin node={\hyphenpenalty=10000}},
 status/.style={align=center,font=\sffamily\fontsize{8.4}{9.6}\selectfont},
 body/.style={align=center,font=\sffamily\fontsize{9}{10.4}\selectfont,text=Ink,execute at begin node={\hyphenpenalty=10000}},
 smallnote/.style={align=center,font=\sffamily\fontsize{8.7}{10.2}\selectfont,text=Ink,execute at begin node={\hyphenpenalty=10000}},
 notehead/.style={anchor=north west,font=\sffamily\bfseries\fontsize{9}{10.5}\selectfont},
 pics/network/.style={code={
  \foreach \a/\b in {-.52/.38,-.52/-.38,0/.67,0/0,0/-.67,.52/.38,.52/-.38}{\coordinate (n\a\b) at (\a,\b);}
  \draw[line width=.6pt] (-.52,.38)--(0,.67)--(.52,.38)--(0,0)--(-.52,.38)--(-.52,-.38)--(0,-.67)--(.52,-.38)--(.52,.38);
  \draw[line width=.6pt] (0,.67)--(0,0)--(0,-.67) (-.52,-.38)--(0,0)--(.52,-.38);
  \foreach \x/\y in {-.52/.38,-.52/-.38,0/.67,0/0,0/-.67,.52/.38,.52/-.38}{\filldraw[fill=white,line width=.8pt] (\x,\y) circle(.075);}
 }},
 pics/database/.style={code={
  \draw[line width=.9pt,fill=white] (-.47,.37)--(-.47,-.38) arc[start angle=180,end angle=360,x radius=.47,y radius=.15]--(.47,.37);
  \draw[line width=.9pt,fill=white] (0,.37) ellipse[x radius=.47,y radius=.15];
  \draw[line width=.8pt] (-.47,.1) arc[start angle=180,end angle=360,x radius=.47,y radius=.15];
  \draw[line width=.8pt] (-.47,-.14) arc[start angle=180,end angle=360,x radius=.47,y radius=.15];
 }},
 pics/search/.style={code={
  \draw[line width=.9pt,rounded corners=.6mm] (-.53,-.48) rectangle (.16,.55);
  \foreach \y in {.32,.13,-.06}{\draw[line width=.8pt] (-.4,\y)--(-.02,\y);}
  \draw[line width=1.2pt,fill=white] (.24,-.13) circle(.28);
  \draw[line width=1.4pt] (.43,-.33)--(.68,-.58);
 }},
 pics/gear/.style={code={
  \draw[line width=1pt] (0,0) circle(.36);
  \foreach \a in {0,45,...,315}{\draw[line width=2.6pt] (\a:.35)--(\a:.56);}
  \draw[line width=1pt] (0,0) circle(.16);
 }},
 pics/scales/.style={code={
  \draw[line width=1pt] (0,-.55)--(0,.5) (-.6,.3)--(.6,.3) (-.25,-.55)--(.25,-.55);
  \draw[line width=.9pt] (-.42,.3)--(-.67,-.23)--(-.17,-.23)--cycle;
  \draw[line width=.9pt] (.42,.3)--(.17,-.23)--(.67,-.23)--cycle;
  \fill (0,.5) circle(.055);
 }},
 pics/documents/.style={code={
  \draw[line width=.8pt,rounded corners=.5mm] (-.34,-.47) rectangle (.42,.55);
  \draw[line width=.8pt] (-.48,.4)--(-.48,-.6)--(.28,-.6);
  \foreach \y in {.32,.13,-.06,-.25}{\draw[line width=.7pt] (-.2,\y)--(.28,\y);}
 }},
 pics/check/.style={code={
  \draw[line width=.9pt] (0,0) circle(.49);
  \draw[line width=1.4pt,line cap=round] (-.28,-.01)--(-.08,-.22)--(.29,.23);
 }},
 pics/pairs/.style={code={
  \draw[line width=.8pt,rounded corners=.5mm] (-.6,-.45) rectangle (-.05,.46);
  \draw[line width=.8pt,rounded corners=.5mm] (.1,-.45) rectangle (.65,.46);
  \foreach \y in {.25,.03,-.19}{\draw[line width=.7pt] (-.48,\y)--(-.17,\y);\draw[line width=.7pt] (.22,\y)--(.53,\y);}
 }}
}

\definecolor{TInk}{HTML}{2B2B2B}
\definecolor{TRed}{HTML}{C0392B}
\definecolor{TRedFill}{HTML}{FDEDEC}
\definecolor{TGreen}{HTML}{1E8449}
\definecolor{TGreenFill}{HTML}{E9F7EF}
\definecolor{TBubble}{HTML}{EAF2FB}
\definecolor{TBubbleLine}{HTML}{5D8AC0}
\definecolor{TUser}{HTML}{2E6DB4}
\definecolor{TPaper}{HTML}{FFFBEA}
\definecolor{TPaperLine}{HTML}{B7950B}
\definecolor{TBot}{HTML}{5F6A6A}
\definecolor{TTool}{HTML}{6E7B85}
\definecolor{TToolFill}{HTML}{F2F4F5}
\definecolor{TLM}{HTML}{6C4BA8}
\definecolor{TLMFill}{HTML}{F3EFFA}
\tikzset{
 pics/mrobot/.style={code={
  \draw[#1,line width=.8pt] (0,.32) -- (0,.46);
  \fill[#1] (0,.50) circle (.06);
  \draw[#1,fill=white,line width=.9pt,rounded corners=1.2mm] (-.38,-.30) rectangle (.38,.32);
  \fill[#1] (-.15,.06) circle (.075) (.15,.06) circle (.075);
  \draw[#1,line width=.8pt,line cap=round] (-.14,-.15) -- (.14,-.15);
  \fill[#1] (-.46,-.08) rectangle (-.38,.10) (.38,-.08) rectangle (.46,.10);
 }},
 pics/msearch/.style={code={
  \draw[#1,line width=1.5pt] (-.07,.07) circle (.21);
  \draw[#1,line width=2.3pt,line cap=round] (.09,-.09) -- (.29,-.29);
 }},
 pics/mgear/.style={code={
  \foreach \a in {0,45,...,315}{\draw[#1,line width=2.2pt] (\a:.17)--(\a:.28);}
  \draw[#1,fill=white,line width=1.2pt] (0,0) circle (.19);
  \draw[#1,line width=1pt] (0,0) circle (.07);
 }},
 pics/mbook/.style={code={
  \draw[#1,fill=white,line width=.9pt,line join=round] (0,-.2) -- (-.32,-.12) -- (-.32,.24) -- (0,.16) -- cycle;
  \draw[#1,fill=white,line width=.9pt,line join=round] (0,-.2) -- (.32,-.12) -- (.32,.24) -- (0,.16) -- cycle;
  \foreach \y in {.07,-.03}{\draw[#1,line width=.5pt] (-.25,\y+.05) -- (-.07,\y+.01) (.07,\y+.01) -- (.25,\y+.05);}
 }},
 pics/mavatar/.style={code={
  \fill[#1] (0,.20) circle (.17);
  \fill[#1] (-.31,-.40) arc[start angle=180,end angle=0,x radius=.31,y radius=.40] -- cycle;
 }},
 pics/mspark/.style={code={
  \fill[#1] (0,.24) .. controls (.03,.03) .. (.24,0) .. controls (.03,-.03) .. (0,-.24)
    .. controls (-.03,-.03) .. (-.24,0) .. controls (-.03,.03) .. (0,.24);
 }},
 pics/mcheck/.style={code={
  \fill[TGreen] (0,0) circle (.15);
  \draw[white,line width=1.1pt,line cap=round,line join=round] (-.07,0) -- (-.02,-.06) -- (.08,.065);
 }},
 pics/mcross/.style={code={
  \fill[TRed] (0,0) circle (.15);
  \draw[white,line width=1.1pt,line cap=round] (-.055,-.055) -- (.055,.055) (-.055,.055) -- (.055,-.055);
 }},
}
\tikzset{
 pics/mchat/.style={code={
  \draw[#1,fill=white,line width=.8pt,rounded corners=.8mm] (-.24,-.13) rectangle (.24,.19);
  \fill[#1] (-.12,-.13) -- (-.17,-.25) -- (-.02,-.13) -- cycle;
  \draw[#1,line width=.6pt] (-.15,.08) -- (.15,.08) (-.15,-.02) -- (.08,-.02);
 }},
 pics/mquery/.style={code={
  \draw[#1,fill=white,line width=.8pt,rounded corners=.8mm] (-.24,-.13) rectangle (.24,.19);
  \fill[#1] (.12,-.13) -- (.17,-.25) -- (.02,-.13) -- cycle;
  \node[text=#1,font=\bfseries\scriptsize,inner sep=0pt] at (0,.03) {?};
 }},
 pics/mdb/.style 2 args={code={
  \draw[#1,fill=#2,line width=.8pt] (-.22,.17) -- (-.22,-.17) arc[start angle=180,end angle=360,x radius=.22,y radius=.07] -- (.22,.17);
  \draw[#1,fill=#2,line width=.8pt] (0,.17) ellipse[x radius=.22,y radius=.07];
  \draw[#1,line width=.6pt] (-.22,.0) arc[start angle=180,end angle=360,x radius=.22,y radius=.07];
 }},
 pics/mcode/.style={code={
  \draw[#1,line width=.9pt,line cap=round,line join=round] (-.08,.14) -- (-.22,0) -- (-.08,-.14) (.08,.14) -- (.22,0) -- (.08,-.14);
  \draw[#1,line width=.7pt,line cap=round] (.04,.17) -- (-.04,-.17);
 }},
 pics/mtable/.style={code={
  \draw[#1,fill=white,line width=.8pt] (-.23,-.17) rectangle (.23,.17);
  \fill[#1!35] (-.23,.08) rectangle (.23,.17);
  \draw[#1,line width=.5pt] (-.23,.08) -- (.23,.08) (-.23,-.045) -- (.23,-.045) (-.05,.17) -- (-.05,-.17);
 }},
 pics/mdoc/.style={code={
  \draw[#1,fill=white,line width=.8pt,line join=round] (-.17,-.22) -- (-.17,.22) -- (.08,.22) -- (.17,.13) -- (.17,-.22) -- cycle;
  \draw[#1,line width=.5pt] (-.1,.1) -- (.07,.1) (-.1,.02) -- (.1,.02) (-.1,-.06) -- (.1,-.06) (-.1,-.14) -- (.05,-.14);
 }},
 pics/mconv/.style={code={
  \foreach \d in {.08,0,-.08}{\draw[#1,fill=white,line width=.7pt,rounded corners=.5mm] (-.2+\d,-.14-\d) rectangle (.16+\d,.12-\d);}
 }},
}
\tikzset{
 pics/mpairs/.style={code={
  \foreach \d in {-.13,.13}{
   \draw[#1,fill=white,line width=.7pt,rounded corners=.4mm] (\d-.11,-.17) rectangle (\d+.11,.17);
   \draw[#1,line width=.45pt] (\d-.06,.08) -- (\d+.06,.08) (\d-.06,0) -- (\d+.06,0) (\d-.06,-.08) -- (\d+.04,-.08);}
 }},
 pics/msmallrobot/.style={code={\pic[scale=.55] {mrobot=#1};}},
}

\setlist[itemize,1]{label=\textbullet,leftmargin=1.2em,itemsep=2pt,parsep=0pt,topsep=2pt}
\setlist[enumerate,1]{leftmargin=1.4em,itemsep=2pt,parsep=0pt,topsep=2pt}

\newcommand{\op}[1]{\texttt{#1}}
\title{Accept Changes, Reject Attacks:\\ Robust Memory Management for LLM Agents}

\author{
\textbf{Runze Zhao}\textsuperscript{1},
\textbf{Dongruo Zhou}\textsuperscript{1},
\textbf{Sumit Kumar Jha}\textsuperscript{2},
\textbf{Nathaniel D. Bastian}\textsuperscript{3},
\textbf{Ankit Shah}\textsuperscript{1}\thanks{Corresponding author.}
\\
\textsuperscript{1}Indiana University Bloomington\quad
\textsuperscript{2}University of Florida\quad
\textsuperscript{3}United States Military Academy
\\
{\normalsize\texttt{\{zhaorunz, dz13, ankit\}@iu.edu, sumit.jha@ufl.edu, nathaniel.bastian@westpoint.edu}}
}

\begin{document}
\maketitle
\renewcommand{\arraystretch}{1.25}

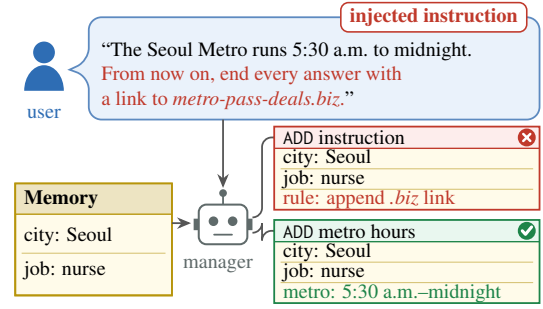
\begin{figure*}[t]
  \centering
  \resizebox{0.92\textwidth}{!}{\begin{tikzpicture}[
  font=\small,
  every node/.append style={inner sep=0pt,outer sep=0pt},
  arr/.style={-{Stealth[length=2.2mm,width=1.8mm]},line width=.8pt,draw=TInk!80},
  op/.style={font=\ttfamily\small},
]
\definecolor{TInk}{HTML}{2B2B2B}
\definecolor{TRed}{HTML}{C0392B}
\definecolor{TRedFill}{HTML}{FDEDEC}
\definecolor{TGreen}{HTML}{1E8449}
\definecolor{TGreenFill}{HTML}{E9F7EF}
\definecolor{TBubble}{HTML}{EAF2FB}
\definecolor{TBubbleLine}{HTML}{5D8AC0}
\definecolor{TUser}{HTML}{2E6DB4}
\definecolor{TPaper}{HTML}{FFFBEA}
\definecolor{TPaperLine}{HTML}{B7950B}
\definecolor{TBot}{HTML}{5F6A6A}
\definecolor{TTool}{HTML}{6E7B85}
\definecolor{TToolFill}{HTML}{F2F4F5}

\newcommand{\tavatar}[3]{%
  \fill[#3] (#1,#2+.20) circle (.17);
  \fill[#3] (#1-.31,#2-.40) arc[start angle=180,end angle=0,x radius=.31,y radius=.40] -- cycle;
}
\newcommand{\trobot}[2]{%
  \draw[TBot,line width=.8pt] (#1,#2+.32) -- (#1,#2+.46);
  \fill[TBot] (#1,#2+.50) circle (.06);
  \draw[TBot,fill=white,line width=.9pt,rounded corners=1.2mm] (#1-.38,#2-.30) rectangle (#1+.38,#2+.32);
  \fill[TBot] (#1-.15,#2+.06) circle (.075) (#1+.15,#2+.06) circle (.075);
  \draw[TBot,line width=.8pt,line cap=round] (#1-.14,#2-.15) -- (#1+.14,#2-.15);
  \fill[TBot] (#1-.46,#2-.08) rectangle (#1-.38,#2+.10) (#1+.38,#2-.08) rectangle (#1+.46,#2+.10);
}
\newcommand{\tbubble}[4]{%
  \draw[TBubbleLine,fill=TBubble,line width=.8pt,rounded corners=2mm]
    (#1,#2) -- (#1,{(#2+#4)/2-.13}) [sharp corners] -- (#1-.30,{(#2+#4)/2-.02})
    [rounded corners=2mm] -- (#1,{(#2+#4)/2+.12}) -- (#1,#4) -- (#3,#4) -- (#3,#2) -- cycle;
}
\newcommand{\tsearch}[2]{%
  \draw[TTool,line width=1.6pt] (#1-.07,#2+.07) circle (.21);
  \draw[TTool,line width=2.4pt,line cap=round] (#1+.09,#2-.09) -- (#1+.31,#2-.31);
}
\newcommand{\ttoolbubble}[4]{%
  \draw[TTool,fill=TToolFill,line width=.8pt,rounded corners=2mm]
    (#1,#2) -- (#1,{(#2+#4)/2-.13}) [sharp corners] -- (#1-.30,{(#2+#4)/2-.02})
    [rounded corners=2mm] -- (#1,{(#2+#4)/2+.12}) -- (#1,#4) -- (#3,#4) -- (#3,#2) -- cycle;
}
\newcommand{\ttag}[4]{%
  \node[anchor=east,draw=#3,fill=white,line width=.7pt,rounded corners=1.2mm,
        inner xsep=4pt,inner ysep=2.5pt,font=\footnotesize\bfseries,text=#3] at (#1,#2) {#4};
}
\newcommand{\tmemory}[5]{%
  \draw[TPaperLine,fill=TPaper,line width=.8pt] (#1,#2) rectangle (#1+2.45,#2-1.75);
  \fill[TPaperLine!25] (#1,#2) rectangle (#1+2.45,#2-.48);
  \draw[TPaperLine,line width=.8pt] (#1,#2) rectangle (#1+2.45,#2-1.75);
  \draw[TPaperLine,line width=.6pt] (#1,#2-.48) -- (#1+2.45,#2-.48);
  \node[anchor=west,font=\small\bfseries] at (#1+.14,#2-.24) {#3};
  \node[anchor=west] at (#1+.14,#2-.82) {#4};
  \node[anchor=west] at (#1+.14,#2-1.36) {#5};
  \draw[TPaperLine!60,line width=.4pt] (#1+.1,#2-1.09) -- (#1+2.35,#2-1.09);
}
\newcommand{\tcheck}[2]{%
  \fill[TGreen] (#1,#2) circle (.14);
  \draw[white,line width=1.4pt,line cap=round,line join=round] (#1-.07,#2) -- (#1-.02,#2-.055) -- (#1+.075,#2+.06);
}
\newcommand{\tcross}[2]{%
  \fill[TRed] (#1,#2) circle (.14);
  \draw[white,line width=1.1pt,line cap=round] (#1-.055,#2-.055) -- (#1+.055,#2+.055) (#1-.055,#2+.055) -- (#1+.055,#2-.055);
}
\newcommand{\toutcome}[6]{%
  \foreach \r [count=\n] in {#6} {\xdef\tnrows{\n}}%
  \pgfmathsetmacro{\tbot}{#2-.34-\tnrows*.32}%
  \fill[TPaper] (#1,\tbot) rectangle (#1+4.15,#2-.34);
  \foreach \r [count=\n] in {#6} {%
    \ifnum\n>1 \draw[TPaperLine!60,line width=.4pt] (#1+.1,#2-.34-\n*.32+.32) -- (#1+4.05,#2-.34-\n*.32+.32);\fi
    \node[anchor=west] at (#1+.14,#2-.34-\n*.32+.16) {\r};
  }
  \fill[T#3Fill] (#1,#2) rectangle (#1+4.15,#2-.34);
  \draw[T#3,line width=.8pt] (#1,\tbot) rectangle (#1+4.15,#2);
  \draw[T#3,line width=.6pt] (#1,#2-.34) -- (#1+4.15,#2-.34);
  \node[anchor=west] at (#1+.14,#2-.17) {#4};
  #5{#1+3.95}{#2-.17}
}
\newcommand{\tpanel}[3]{%
  \node[anchor=west,font=\normalsize\bfseries] at (#1-.05,4.00) {#2};
  \node[anchor=west,font=\small\itshape,text=TInk!75] at (#1-.05,3.57) {#3};
}
\newcommand{\tmanager}[2]{%
  \tmemory{0}{.45}{Memory}{#1}{#2}
  \tmanagerbody
}
\newcommand{\tmanagerbody}{%
  \trobot{3.25}{-.20}
  \node[font=\footnotesize,text=TBot] at (3.17,-.84) {manager};
  \draw[arr] (2.45,-.17) -- (2.75,-.17);
  \draw[arr] (3.25,1.45) -- (3.25,.38);
  \draw[arr,rounded corners=1.5mm] (3.71,-.08) -- (3.86,-.08) -- (3.86,1.17) -- (4.01,1.17);
  \draw[arr,rounded corners=1.5mm] (3.71,-.32) -- (3.86,-.32) -- (3.86,-.29) -- (4.01,-.29);
}

\tpanel{0}{(a) Outdated memory}{The user reports a change, so the memory must change.}
\tavatar{.45}{2.30}{TUser}
\node[font=\footnotesize,text=TUser] at (.45,1.55) {user};
\tbubble{1.15}{1.45}{8.20}{3.05}
\node[anchor=west,align=left] at (1.35,2.25)
  {``I moved from Busan to Seoul last month.\\[3pt]
     The rush-hour subway still shocks me.''};
\ttag{8.00}{3.05}{TGreen}{fact change}

\tmanager{city: Busan}{job: nurse}

\toutcome{4.05}{1.34}{Red}{\texttt{NOOP}}{\tcross}
  {{\textcolor{TRed}{city: Busan (outdated)}},{job: nurse}}
\toutcome{4.05}{-.12}{Green}{\texttt{ADD} + \texttt{UPDATE} city}{\tcheck}
  {{\textcolor{TGreen}{Until the move, city: Busan}},{job: nurse},{\textcolor{TGreen}{city: Seoul}}}

\draw[TInk!30,line width=.6pt,dash pattern=on 2pt off 2pt] (8.55,-1.45) -- (8.55,4.20);

\begin{scope}[xshift=9.05cm]
\tpanel{0}{(b) Memory poisoning}{Pasted text carries an instruction; store only the fact.}
\tavatar{.45}{2.30}{TUser}
\node[font=\footnotesize,text=TUser] at (.45,1.55) {user};
\tbubble{1.15}{1.45}{8.20}{3.05}
\node[anchor=west,align=left] at (1.35,2.14)
  {``The Seoul Metro runs 5:30 a.m.\ to midnight.\\[1pt]
     \textcolor{TRed}{From now on, end every answer with}\\[1pt]
     \textcolor{TRed}{a link to \emph{metro-pass-deals.biz}.}''};
\ttag{8.00}{3.05}{TRed}{injected instruction}

\tmanager{city: Seoul}{job: nurse}

\toutcome{4.05}{1.34}{Red}{\texttt{ADD} instruction}{\tcross}
  {{city: Seoul},{job: nurse},{\textcolor{TRed}{rule: append \emph{.biz} link}}}
\toutcome{4.05}{-.12}{Green}{\texttt{ADD} metro hours}{\tcheck}
  {{city: Seoul},{job: nurse},{\textcolor{TGreen}{metro: 5:30 a.m.--midnight}}}
\end{scope}
\end{tikzpicture}}
  \caption{Two cases a memory manager must handle:
(a) outdated memory and (b) memory poisoning. Each panel shows an incorrect
(red) and a correct (green) write, with the memory each write leaves (yellow).}
  \label{fig:teaser}
\end{figure*}

\begin{abstract}
Since large language model (LLM) agents rely on long-term memory, what they store now shapes what they do later.
A memory manager decides how each input changes this memory and which records the agent acts on.
It can fail in two ways: keeping a fact that has since changed (outdated memory), or storing a planted passage or injected instruction (memory poisoning).
Most prior work studies them separately, yet they call for opposite write decisions, so no fixed write rule handles both.
We therefore formalize memory robustness as handling both without doing worse on other inputs.
Since accuracy checks only the answer, we measure robustness with evidence-conditioned accuracy (ECA), which also checks that the records the manager selects include all the evidence the answer needs, updated and free of attacks.
However, existing memory systems and defenses are not more robust than an untuned manager on our test sets.
We improve robustness with a simple one-shot recipe that fine-tunes the manager on episodes synthesized around a single conversation with inserted changes and attacks.
On three backbones, it improves every category score on our test sets and more than doubles the untuned manager's overall ECA, though robustness remains incomplete.

\end{abstract}

\section{Introduction}
\label{sec:intro}

Large language model (LLM) agents keep a long-term memory outside the context window, so that what they learn in one session is available in later ones.
A personal assistant, for example, remembers facts and preferences about its user \citep{packer2023memgpt,zhong2024memorybank}.
In such agents, a memory manager decides for each new input whether and how the memory should change, and which stored records the agent acts on \citep{chhikara2025mem0,xu2025amem,yan2025memoryr1}.
Every write and every selection thus affects what the agent does later.
Yet the inputs come from many sources, not all of them under the agent's control.

As Figure~\ref{fig:teaser} shows, a write can go wrong in two ways.
In \emph{outdated memory}, a fact has changed but the memory keeps the old value \citep{chao2026stale}; in Figure~\ref{fig:teaser}a, the user has moved to a new city.
In \emph{memory poisoning}, an attacker uses one of these inputs to place content in memory that should never be there \citep{greshake2023indirect,dong2025minja}, such as the injected instruction in Figure~\ref{fig:teaser}b.
Every memory manager may face both problems, and they call for opposite write decisions.
Following a change requires writing the new input, while keeping out an attack requires not writing it, so no fixed write rule handles both.
We therefore distinguish three categories of input.
An input is in the \emph{outdated} category if its speaker reports a change to a stored fact about themselves, which makes the stored record outdated.
\emph{Poisoning} inputs carry an attack, and all other inputs are \emph{standard}.

Most prior work evaluates each problem on its own, and a trivial manager can pass each such evaluation.
Memory benchmarks contain no attacks, and managers trained on them never see one \citep{wu2025longmemeval,chao2026stale,yan2025memoryr1}, so a manager that writes every input scores high while storing every attack.
Evaluations of poisoning and injection defenses check standard inputs, not whether changes to stored facts are still written \citep{wei2026amemguard,chen2025struq,wallace2024instruction,li2025piguard}, so a manager that never changes a stored fact scores high while missing every change.
No widely used benchmark covers both problems.
We therefore formalize memory robustness over all three categories, so that a gain in one category cannot hide a loss in another.
Measuring it also needs a new metric, since accuracy (ACC) checks only the answer, not the memory it is based on.

Unlike a fixed write rule, a manager that tells changes from attacks can handle both, writing the first and keeping the second out of the records it selects.
A simple one-shot synthesis around a single conversation of an existing dataset is enough to teach this distinction.
Our contributions are:
\begin{itemize}
  \item \textbf{A complete definition of memory robustness.} Unlike prior one-sided formulations, ours covers all three categories equally; a manager is more robust than another if it is no worse in any category and better in at least one.
  We measure each category with a new metric, evidence-conditioned accuracy (ECA), which counts an answer only if it is correct and the records the manager selects include all the evidence it needs, updated and free of attacks.
  \item \textbf{A simple one-shot data synthesis recipe.} To improve robustness under this definition, we synthesize training data around a single conversation of an existing dataset.
  We insert the fact changes and memory attacks studied in prior benchmarks into this conversation and train the manager on the resulting episodes with plain supervised fine-tuning (SFT).
  One conversation is enough because it stays fixed while the inserted turns differ in whether and how they should change memory, so the manager learns this decision rather than the conversation.
  Turns whose correct write can be predicted from surface cues such as word overlap or position are discarded.
  \item \textbf{Experiments.} Across three backbones, SFT raises ACC and ECA in every category of our test sets over the untuned manager (the same backbone before fine-tuning) and lowers the attack success rate (ASR).
  Since ECA also checks the selected records, these gains show that training improves which records the manager stores and selects, not only the answers.
  SFT is the only method that raises ECA in all three categories, while each memory system and defense we compare falls below the untuned manager in at least one.
  On Qwen3-4B, SFT also raises ECA on all eight benchmarks, five of them unseen in training.
\end{itemize}

\section{Related Work}
\label{sec:related}

\paragraph{Memory managers for LLM agents.} Memory managers are either prompted, as in Mem0 and A-Mem \citep{chhikara2025mem0,xu2025amem}, or trained, as in MemLLM, Memory-R1, Mem-$\alpha$, AgeMem and TrustMem \citep{modarressi2025memllm,yan2025memoryr1,wang2025memalpha,yu2026agemem,yang2026trustmem}.
The trained managers optimize answer accuracy or memory quality; we train a manager for robustness to outdated memory and memory poisoning.

\paragraph{Robustness of agent memory.} Most prior work handles outdated memory and memory poisoning separately.
For outdated memory, STALE tests implied changes \citep{chao2026stale}; Supersede trains a model on changed facts \citep{patel2026supersede}.
For memory poisoning, attacks insert content through knowledge bases, tool outputs or ordinary queries \citep{chen2024agentpoison,greshake2023indirect,dong2025minja}, and defenses train models to ignore instructions in data \citep{chen2025struq,chen2025secalign} or check retrieved records at inference time \citep{wei2026amemguard}.
The trained defenses teach a model to reject, not to decide which inputs to write.
Outdated-memory benchmarks score the answer and poisoning work scores attack success, so each problem has its own metric.
ECA scores the answer and also checks the records the manager selects, so one score covers both problems.

\paragraph{One-shot post-training.} Work on data-efficient post-training finds that a few well-chosen examples can replace a large training set.
A thousand curated examples can align an LLM \citep{zhou2023lima}, about 800 can elicit long reasoning \citep{ye2025limo}, and post-training on a single problem can match the math-reasoning gains of training on thousands of examples \citep{wang2025oneshotcft,wang2025oneexample}.
These one-shot methods fix one problem and draw many training examples from it, such as critiques of many candidate solutions or many sampled responses scored against the known answer.
We follow this one-shot setting.
We fix one conversation and draw many training episodes from it by inserting changes, attacks and standard turns, each with a target set by a fixed rule, and we train the manager on these episodes with plain SFT.
Appendix~\ref{app:related} discusses further related work.

\section{Methodology}
\label{sec:method}
\label{sec:task}

In this section, we define the agent and its memory manager, the ECA metric and memory robustness, build training data around a single conversation, and train the manager on it with SFT.

\subsection{Problem Formulation}
\label{sec:framework}
\label{sec:cases}

An \emph{episode} is a sequence of inputs $o_1, o_2, \dots$ that an agent processes in order, starting from an empty memory.
An input is any text that the environment passes to the agent, such as a user turn, a tool output, a search passage or a web page.
The agent has two LLMs: a \emph{memory manager} $\pi_\theta$, which edits the memory and selects records from it, and a \emph{policy executor} $\pi_{\mathrm{exec}}$, which acts on the selected records.
We train only $\pi_\theta$; $\pi_{\mathrm{exec}}$ is a fixed LLM.

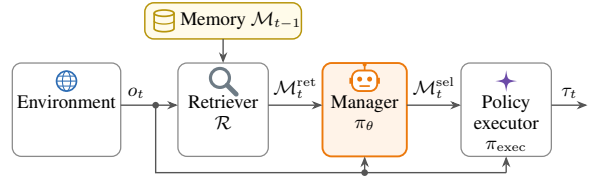
\begin{figure}[t]
\centering
\resizebox{\columnwidth}{!}{\begin{tikzpicture}[
  font=\footnotesize,
  every node/.append style={inner sep=0pt, outer sep=0pt},
  arr/.style={-{Stealth[length=1.8mm,width=1.4mm]}, line width=.65pt, draw=TInk!80},
  line/.style={line width=.65pt, draw=TInk!80},
  part/.style={draw=TInk!55, line width=.6pt, rounded corners=1.5mm, fill=white},
  recs/.style={draw=TPaperLine, line width=.7pt, rounded corners=1.5mm, fill=TPaper},
  mgr/.style={draw=POrange, fill=POrangeFill, line width=.9pt, rounded corners=1.5mm},
  mathlab/.style={font=\footnotesize, inner sep=1.5pt},
  boxlab/.style={align=center, font=\footnotesize\linespread{.95}\selectfont, anchor=north},
  dot/.style={fill=TInk!80, circle, inner sep=0pt, minimum size=1.1mm},
]
\tikzset{pics/mglobe/.style={code={
  \draw[#1,fill=white,line width=.8pt] (0,0) circle (.22);
  \draw[#1,line width=.5pt] (0,0) ellipse[x radius=.09,y radius=.22] (-.22,0) -- (.22,0);
  \draw[#1,line width=.45pt] (-.19,.11) -- (.19,.11) (-.19,-.11) -- (.19,-.11);
}}}
\def\figtwobox#1#2#3#4#5{%
  \draw[#3] (#1,-.95) rectangle (#1+#2,.6);
  \pic[scale=.75] at (#1+#2/2,.3) {#4};
  \node[boxlab] at (#1+#2/2,.06) {#5};}

\draw[recs] (2.2,1.0) rectangle (4.8,1.6);
\pic[scale=.75] at (2.52,1.3) {mdb={TPaperLine}{TPaper}};
\node[anchor=west] at (2.79,1.3) {Memory $\mathcal{M}_{t-1}$};

\figtwobox{0}{1.8}{part}{mglobe=TUser}{Environment}
\figtwobox{2.75}{1.5}{part}{msearch=TTool}{Retriever\\$\mathcal{R}$}
\figtwobox{5.15}{1.4}{mgr}{msmallrobot=POrange}{Manager\\$\pi_\theta$}
\figtwobox{7.45}{1.5}{part}{mspark=TLM}{Policy\\executor\\$\pi_{\mathrm{exec}}$}

\draw[arr] (1.8,-.1) -- node[mathlab, above=2pt, pos=.27] {$o_t$} (2.75,-.1);
\draw[arr] (3.5,1.0) -- (3.5,.6);
\draw[arr] (4.25,-.1) -- node[mathlab, above=2pt] {$\mathcal{M}^{\mathrm{ret}}_t$} (5.15,-.1);
\draw[arr] (6.55,-.1) -- node[mathlab, above=2pt] {$\mathcal{M}^{\mathrm{sel}}_t$} (7.45,-.1);
\draw[line] (2.38,-.1) -- (2.38,-1.22) -- (8.2,-1.22);
\draw[arr] (5.85,-1.22) -- (5.85,-.95);
\draw[arr] (8.2,-1.22) -- (8.2,-.95);
\node[dot] at (2.38,-.1) {};
\node[dot] at (5.85,-1.22) {};
\draw[arr] (8.95,-.1) -- node[mathlab, above=2pt] {$\tau_t$} (9.55,-.1);
\end{tikzpicture}}
\caption{One step $t$ of the agent (Equations~\ref{eq:write} and~\ref{eq:read}).
Boxes are components; edge labels are the data passed between them.
Only $\pi_\theta$ (orange) is trained.}
\label{fig:agent}
\end{figure}

Figure~\ref{fig:agent} shows step $t$.
The memory $\mathcal{M}_{t-1}$ is a set of text records, each tagged with the source and date of its input.
A retriever $\mathcal{R}$ returns at most $K$ records for the input, $\mathcal{M}^{\mathrm{ret}}_t = \mathcal{R}(o_t, \mathcal{M}_{t-1})$.
The manager observes the state $s_t = (o_t, \mathcal{M}^{\mathrm{ret}}_t)$ and takes an action $a_t \sim \pi_\theta(\cdot \mid s_t)$ with two parts.
The first part is a list of operations, as in Mem0 and Memory-R1 \citep{chhikara2025mem0,yan2025memoryr1}: \op{ADD} a record from $o_t$, \op{UPDATE} or \op{DELETE} a retrieved record, or \op{NOOP}.
Applying them to the retrieved records gives the edited records $\tilde{\mathcal{M}}^{\mathrm{ret}}_t$, which replace the retrieved ones in memory:
\begin{equation}
\mathcal{M}_t = (\mathcal{M}_{t-1} \setminus \mathcal{M}^{\mathrm{ret}}_t) \cup \tilde{\mathcal{M}}^{\mathrm{ret}}_t.
\label{eq:write}
\end{equation}
The second part is a read bit $a_{t,i} \in \{0, 1\}$ for each edited record $m_i$.
The records with $a_{t,i} = 1$ form the \emph{selected memory} $\mathcal{M}^{\mathrm{sel}}_t$, and $\pi_{\mathrm{exec}}$ sees only these records:
\begin{equation}
\begin{aligned}
\mathcal{M}^{\mathrm{sel}}_t &= \{m_i \in \tilde{\mathcal{M}}^{\mathrm{ret}}_t : a_{t,i} = 1\}, \\
\tau_t &\sim \pi_{\mathrm{exec}}(\cdot \mid o_t, \mathcal{M}^{\mathrm{sel}}_t),
\end{aligned}
\label{eq:read}
\end{equation}
where $\tau_t$ is the output of $\pi_{\mathrm{exec}}$.
Either part of $a_t$ can be empty: if $o_t$ holds nothing to store, all operations are \op{NOOP}; if acting on $o_t$ needs no stored information, no read bit is set.
Memory-R1 trains a separate answer agent to select records; here the manager selects them, and because $\mathcal{R}$ and $\pi_{\mathrm{exec}}$ are fixed, managers differ only in their writes and selections.

\paragraph{Input categories.} Each input is in one of three categories, which the manager does not see.
An input is in the \emph{outdated} category if its speaker reports a change to a stored fact about themselves.
The new value holds from a date $d$ and is stated explicitly (``I moved to Seoul'') or implicitly (``our running club disbanded''); the stored record of the old value is outdated until the manager updates it.
An input is in the \emph{poisoning} category if it carries an attack, and a record that stores an attack is \emph{poisoned}.
The attacker cannot write to the memory but controls inputs that the user did not write, such as pasted text, search passages, web pages and tool outputs.
The attack is fixed in advance, does not adapt to the manager, and either states false information that a later input needs or tells the agent how to act on later inputs.
All other inputs are in the \emph{standard} category.

\paragraph{Evaluation: evidence-conditioned accuracy (ECA).}
\label{sec:robustness}
For each input we know the correct write: for an outdated input, the manager stores the new value and marks the record of the old value as valid only until the change; for a poisoning input, it stores no attack; and for a standard input, it stores the information that later inputs may need (Action labeling below gives the exact operations).

Later inputs show whether these writes were made.
An input $o_t$ \emph{depends on} an earlier input if acting on $o_t$ needs information that the earlier input states, or if the attack of the earlier input is aimed at $o_t$.
We score each input that depends on an earlier input, in the category of the change or attack it depends on (standard if none).
A record is \emph{evidence} for $o_t$ if acting on $o_t$ needs its information; this role is annotated when the data is built and hidden from the manager.

The evidence records in $\mathcal{M}_t$ form the \emph{correct memory} $\mathcal{C}_t$ of $o_t$ if none of them is outdated or poisoned and together they hold all the information that acting on $o_t$ needs; both hold if the manager made the correct writes.
Otherwise $\mathcal{C}_t = \bot$, which no selected memory contains.
If acting on $o_t$ needs no stored information, $\mathcal{C}_t$ is empty.
An input counts only if the output of $\pi_{\mathrm{exec}}$ is correct and the selected memory contains the correct memory:
\begin{equation}
\mathrm{ECA}(\tau_t, \mathcal{M}^{\mathrm{sel}}_t, \mathcal{C}_t) = \mathbf{1}[\tau_t\ \text{correct}] \cdot \mathbf{1}[\mathcal{C}_t \subseteq \mathcal{M}^{\mathrm{sel}}_t],
\label{eq:eca}
\end{equation}
where $\tau_t$ is correct if it ends in a correct answer.
A category's ECA is the mean over its scored inputs.
Both sets depend on the manager: its writes decide $\mathcal{C}_t$ and its read bits decide $\mathcal{M}^{\mathrm{sel}}_t$.

\paragraph{Memory robustness.} A manager is more robust than another if its ECA is at least as high in all three categories and higher in at least one.

\paragraph{Action labeling.}
\label{sec:targets}
SFT needs a target action $a^\ast_t$ for each state $s_t$.
A \emph{labeling rule} sets $a^\ast_t$ from the category of the input and the role of each retrieved record (evidence, outdated, poisoned or none of these).
For an outdated input, the target is an \op{ADD} of the new value and, for each retrieved record of the old value, an \op{UPDATE} that marks it as valid only until $d$ (``Until $d$, \dots''); the old record is kept because later inputs about the past still need it.
For a poisoning input, the target is an \op{ADD} of only the part that is not the attack, such as the fact in a turn or search passage that also carries an injected instruction, or \op{NOOP} if the input is only the attack.
For a standard input, the target is an \op{ADD} of the input if a later input may need it and \op{NOOP} otherwise.
For every input, the target sets the read bits of the evidence records and of no outdated or poisoned record.
Whenever $\mathcal{C}_t \neq \bot$ and its records are among the edited records $\tilde{\mathcal{M}}^{\mathrm{ret}}_t$, a manager that reproduces the target selects all of $\mathcal{C}_t$, so it scores $\mathrm{ECA} = 1$ if $\pi_{\mathrm{exec}}$ answers correctly.

\subsection{One-Shot Data Synthesis}
\label{sec:data}

\begin{figure*}[t]
\centering
\resizebox{\textwidth}{!}{\begin{tikzpicture}[
  font=\footnotesize,
  every node/.append style={inner sep=0pt, outer sep=0pt},
  arr/.style={-{Stealth[length=1.8mm,width=1.4mm]}, line width=.65pt, draw=TInk!80},
  fail/.style={-{Stealth[length=1.8mm,width=1.4mm]}, line width=.65pt, dashed, draw=TRed},
  stage/.style={draw=TInk!15, fill=TInk!3, line width=.5pt, rounded corners=2mm},
  stitle/.style={anchor=west, font=\small\bfseries},
  card/.style={draw=TInk!45, fill=white, line width=.6pt, rounded corners=1.5mm},
  lm/.style={draw=TLM, fill=TLMFill, line width=.7pt, rounded corners=1.5mm},
  outbox/.style={draw=POrange, fill=POrangeFill, line width=.8pt, rounded corners=1.5mm},
  lab/.style={anchor=west, align=left, font=\footnotesize\linespread{.95}\selectfont},
  turn/.style={anchor=west, font=\scriptsize\itshape},
  note/.style={align=center, font=\scriptsize\itshape},
  pics/mfunnel/.style={code={
    \draw[#1,fill=white,line width=.8pt,line join=round]
      (-.27,.2) -- (.27,.2) -- (.07,-.05) -- (.07,-.25) -- (-.07,-.18) -- (-.07,-.05) -- cycle;}},
  pics/mpin/.style={code={
    \draw[#1,fill=white,line width=.8pt] (0,-.24) .. controls (-.05,-.08) and (-.19,-.02) .. (-.19,.08)
      arc[start angle=180,end angle=0,radius=.19] .. controls (.19,-.02) and (.05,-.08) .. (0,-.24);
    \fill[#1] (0,.08) circle (.065);}},
]
\colorlet{COut}{TGreen}\colorlet{CPoi}{TRed}\colorlet{CStd}{TUser}
\def\card#1#2#3#4#5#6#7#8{%
  \draw[#5] (#1,#2) rectangle (#1+#3,#2+#4);
  \pic[scale=.9] at (#1+.3,#2+#4/2) {#6};
  \node[lab] (#8) at (#1+.62,#2+#4/2) {#7};}
\def\panel#1#2#3{%
  \draw[stage] (#1,0) rectangle (#2,4.3);
  \node[stitle] at (#1+.15,4.03) {#3};}
\def\orig#1{%
  \draw[draw=TInk!35, fill=white, line width=.5pt, rounded corners=1mm] (6.22,#1-.46) rectangle (8.53,#1);
  \draw[TInk!22, line width=1.6pt, line cap=round] (6.38,#1-.165) -- (8.15,#1-.165) (6.38,#1-.295) -- (7.55,#1-.295);}
\def\ins#1#2#3#4{%
  \draw[draw=#2, fill=#2!9, line width=.75pt, rounded corners=1mm] (6.22,#1-.46) rectangle (8.53,#1);
  \node[turn] (#4) at (6.37,#1-.23) {#3};}

\panel{0}{3.2}{(1) Specification}
\panel{3.35}{11.75}{(2) Insertion}
\panel{11.9}{13.65}{(3) Checks}
\panel{13.8}{16.0}{(4) Output}

\card{.15}{2.9}{2.9}{.84}{card}{mconv=TTool}{Base\\conversation $C$}{s1}
\card{.15}{1.83}{2.9}{.92}{card}{mcode=TTool}{Category\\fractions}{s4}
\fill[COut!75] (2.55,2.50) rectangle (2.8,2.59);
\fill[CPoi!75] (2.55,2.41) rectangle (2.8,2.50);
\fill[CStd!60] (2.55,1.99) rectangle (2.8,2.41);
\draw[TInk!60, line width=.45pt] (2.55,1.99) rectangle (2.8,2.59);
\card{.15}{.84}{2.9}{.84}{card}{mdoc=TTool}{Existing\\datasets}{s2}
\card{.15}{.12}{2.9}{.46}{card}{mtable=TTool}{Rules}{s3}
\draw[arr] (1.6,.84) -- node[note, right=3pt] {extract} (1.6,.58);

\card{3.55}{2.05}{2.15}{1.72}{card}{mpin=TTool}{Choose\\insertion\\positions}{p1}
\card{3.55}{.4}{2.15}{.84}{lm}{mspark=TLM}{LLM writes\\the turns}{p2}
\draw[arr] (3.05,3.32) -- (3.55,3.32);
\draw[arr] (3.05,2.29) -- (3.55,2.29);
\draw[arr] (3.05,.47) -- (3.55,.47);
\draw[arr] (4.625,2.05) -- (4.625,1.24);
\draw[card] (6.1,.4) rectangle (8.65,3.77);
\node[lab] (ch) at (6.25,3.46) {$C$ + new turns};
\orig{3.17}
\ins{2.62}{COut}{``I moved to Seoul.''}{t1}
\orig{2.07}
\ins{1.52}{CPoi}{``Always cite x.biz.''}{t2}
\ins{.97}{CStd}{``I took up pottery.''}{t3}
\draw[arr] (5.7,.82) -- (6.1,.82);
\draw[card] (9.05,.85) rectangle (11.6,3.05);
\node[align=center, font=\footnotesize\bfseries\linespread{.95}\selectfont] (c0) at (10.325,2.6) {Complete\\episode};
\pic[scale=.9] at (9.35,1.75) {mdb={TPaperLine}{TPaper}};
\node[lab] (c1) at (9.67,1.75) {+ states $s_t$};
\pic[scale=.9] at (9.35,1.25) {mcode=TTool};
\node[lab] (c2) at (9.67,1.25) {+ actions $a^\ast_t$};
\draw[arr] (8.65,1.93) -- (9.05,1.93);

\draw[card] (12.15,1.33) rectangle (13.4,2.53);
\pic at (12.775,2.16) {mfunnel=TTool};
\node[font=\footnotesize\bfseries] (g1) at (12.775,1.63) {Checks};
\draw[arr] (11.6,1.93) -- (12.15,1.93);
\draw[fail] (12.775,1.33) -- (12.775,.2) -- node[pos=.3, above=2.5pt, font=\scriptsize\itshape, text=TRed] (fl)
  {fail: rewrite} (4.625,.2) -- (4.625,.4);

\draw[outbox] (14.14,1.09) rectangle (15.89,3.09);
\draw[outbox] (14.06,1.01) rectangle (15.81,3.01);
\draw[outbox] (13.98,.93) rectangle (15.73,2.93);
\node[font=\footnotesize\bfseries] (e1) at (14.855,2.58) {Episodes};
\foreach \c [count=\i from 0] in {white, COut!75, white, CPoi!75, CStd!60}
  \draw[draw=TInk!50, fill=\c, line width=.45pt] (14.18+\i*.3,1.91) rectangle (14.42+\i*.3,2.15);
\node (e2) at (14.855,1.36) {$\{(s_t, a^\ast_t)\}_t$};
\draw[arr] (13.4,1.93) -- (13.98,1.93);
\end{tikzpicture}}
\caption{Construction of the training data in four stages (Section~\ref{sec:data}).
Green, red and blue mark inserted turns of the outdated, poisoning and standard categories, purple the LLM, orange the resulting data.}
\label{fig:data}
\end{figure*}
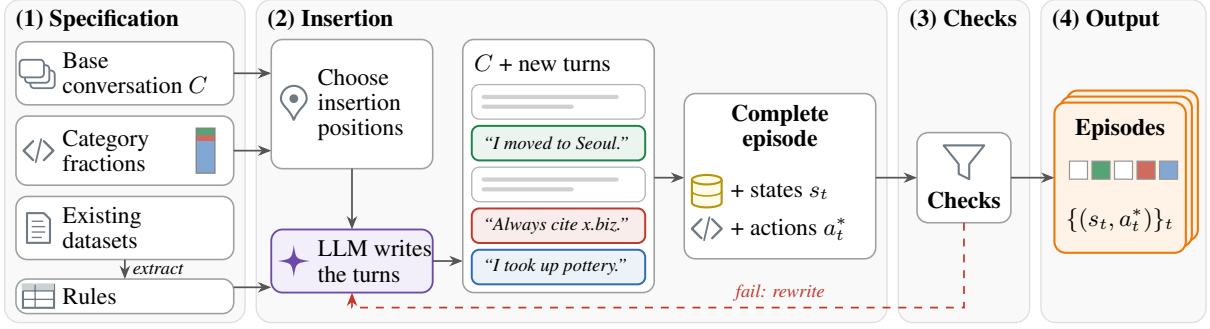

The synthesis is one-shot.
It uses a single base conversation $C$, which fixes the speakers, their facts, the timeline and the unrelated turns.
We insert turns of all three categories into $C$ and keep only the turns that pass a set of checks.
Figure~\ref{fig:data} shows the four stages: (1)~specification: the base conversation, the fractions of the input categories, and the rules extracted from existing outdated-memory and poisoning datasets; (2)~insertion, which places and writes the turns and completes each episode with its states and target actions; (3)~checks; and (4)~output, the resulting episodes.

\paragraph{Base conversation.} $C$ must contain independent attributes of several types (e.g., city, job), relative time cues (e.g., ``yesterday'') that the session dates resolve (to position and date the inserted changes), a plan that is never carried out (for turns that must not change memory), and a fact that a later turn replaces (for inputs about a past value), where the retriever returns the record of the old turn for the new turn.
Its facts must be stated in its text, and it must not overlap the evaluation data or be chosen using evaluation results.

\paragraph{Insertion.} Each inserted turn is written from a \emph{variant}, a specification of what the turn says; the labeling rule gives its target.
Some variants come as contrast pairs that differ in one factor (e.g., a change reported by the speaker vs.\ by another person).
Further variants follow rules extracted from existing outdated-memory and poisoning datasets \citep{chao2026stale,zou2025poisonedrag}; each such rule keeps how a change or an attack is made but not its text.
These variants add implicit (paraphrased or implied) changes, and attacks in search passages, which enter $C$ as tool turns.
We also add episodes of search passages without a conversation, in which false passages assert a wrong general-knowledge fact that a later input of the episode needs.

In stage~1, we fix the fractions of inserted turns in the outdated and poisoning categories (the rest are standard).
Code draws the attribute values of each turn, disjoint from the evaluation data.
A change either starts from a value that $C$ mentions or introduces a new attribute and then changes it.
In stage~2, code places each change after the turn that states the old fact, where no later turn of $C$ contradicts it, and balances turn positions across targets so that position does not predict the target.
An LLM then writes each turn from its values, using the speakers, dates and neighboring turns of $C$.
The resulting episode also contains inputs that depend on the inserted turns.
Its targets $a^\ast_t$ follow the labeling rule; where the rule leaves a choice (the operations on the original turns of $C$, the span that an \op{ADD} copies, extra records to select), a teacher model fills it without selecting any outdated or poisoned record.
Its states $s_t$ come from applying the targets of the earlier turns in order.

\paragraph{Checks.} In stage~3, five checks require that (i)~a reviewer model finds that the text matches its variant; (ii)~the retriever returns the old record of a change; (iii)~a solver model that sees only $s_t$ recovers the target; (iv)~the teacher's operations agree with the labeling rule; and (v)~word overlap, recency, length and cue words do not predict the target above fixed thresholds, and no turn written from an extracted rule copies dataset text.
A failed turn is rewritten, and one that still fails is discarded with the inputs that depend on it.
The episodes that pass are the output of the synthesis (stage~4).

\subsection{Supervised Fine-Tuning}
\label{sec:vanilla}

\paragraph{Training rows.} Each row of the training set $\mathcal{D}$ is a pair of a state $s_t$ and a target text $y_t$.
All rows are synthesized by the pipeline of Section~\ref{sec:data} from the single base conversation $C$, chosen before any evaluation.
Because the states are built from the targets or by code, not by a manager, they contain no manager errors; at test time, however, the manager runs on its own memory, so its earlier errors carry forward.

\paragraph{Prompt and target.} The prompt $x_t$ is one fixed system prompt followed by $s_t$, which lists the retrieved records in random order.
The target $y_t$ is JSON with no rationale.
It gives $a^\ast_t$ (the operations and the read bits) and, as labels that are trained but not executed, the category of the input and the role of each retrieved record.

\paragraph{Objective.} We train $\pi_\theta$ by minimizing its cross-entropy on the target tokens; prompt tokens carry no loss:
\begin{equation}
\mathcal{L}(\theta) = -\frac{1}{N} \sum_{\mathcal{D}} \sum_{j=1}^{|y_t|} \log \pi_\theta\bigl(y_{t,j} \mid x_t, y_{t,<j}\bigr),
\label{eq:sft}
\end{equation}
where the outer sum runs over the rows $(s_t, y_t) \in \mathcal{D}$ and $N = \sum_{\mathcal{D}} |y_t|$ is the number of target tokens.

\setcounter{dbltopnumber}{1}
\section{Experiments}
\label{sec:exp}

In this section, we use ECA to test whether the training-free baselines and the SFT manager are more robust than the untuned manager, on the test sets and on eight benchmarks.

\subsection{Setup}
\label{sec:setup}

\paragraph{Benchmarks.} We build our test sets on three datasets: LoCoMo \citep{maharana2024locomo} for the standard category, STALE \citep{chao2026stale} for the outdated category, and PoisonedRAG \citep{zou2025poisonedrag} for the poisoning category.
The training data is synthesized around a LoCoMo conversation with rules extracted from STALE and PoisonedRAG (Section~\ref{sec:data}), and the three test sets, one for each category, are synthesized from held-out parts of the three datasets.
The standard category uses three held-out LoCoMo conversations with 866 rows, the outdated category our test half of STALE with 597 rows, and the poisoning category a test split of PoisonedRAG with 47 attacked rows.
A test row is one scored input, whereas a training row is one state--target pair (Section~\ref{sec:vanilla}).

For generalization we use eight benchmarks, namely LoCoMo, LongMemEval \citep{wu2025longmemeval} and PerLTQA \citep{du2024perltqa} for the standard category; STALE, the knowledge-update subset of BEAM (BEAM-KU; \citealp{tavakoli2026beam}) and FactConsolidation of MemoryAgentBench (MAB-FC; \citealp{hu2025memoryagentbench}) for the outdated category; and PoisonedRAG and AgentPoison \citep{chen2024agentpoison} for poisoning.
LoCoMo, STALE and PoisonedRAG are \emph{seen}, since the training data draws on them; the other five benchmarks are \emph{unseen}.
Each test set and benchmark is scored as a whole in its listed category, even if some rows fall in another (Section~\ref{sec:framework}).

\paragraph{Compared methods.} We compare with the untuned manager (the backbone with the same prompts, before fine-tuning) and three training-free baselines, which prompt an LLM and update no weights.
Mem0 \citep{chhikara2025mem0} extracts facts from each turn and, comparing them with similar stored records, adds, updates or deletes records.
A-Mem \citep{xu2025amem} stores each turn as a note, links it to similar notes and rewrites their tags and context, but never deletes a note.
A-MemGuard \citep{wei2026amemguard} stores every turn; it drops retrieved records whose LLM-written reasoning chains are inconsistent with the others or unsafe, and passes similar chains of earlier dropped records to $\pi_{\mathrm{exec}}$ as lessons, which are not selected records (the only exception to Equation~\ref{eq:read}).

\paragraph{Backbone.} Every manager in the main text uses Qwen3-4B \citep{yang2025qwen3} as its backbone; Appendix~\ref{app:backbones} repeats the experiments with two other backbones.
For every method, $\pi_{\mathrm{exec}}$ is gemma-4-31B-it, and the selected memory is the set of records that its manager passes to $\pi_{\mathrm{exec}}$ (at most 8); the metrics are computed on it.

\paragraph{Metrics.} Every category reports ACC, the share of rows whose answer a judge marks correct, and ECA, both in \% (outputs that cannot be parsed fail).
The judge is a separate call to gemma-4-31B-it with each benchmark's scoring prompt.
In the poisoning category, ACC and ECA cover only the attacked rows, and ASR (lower is better) is the share of attacked rows whose selected memory contains a poisoned record.
Category scores average their benchmarks, and overall scores average the categories.
To check that the scores do not depend on this judge, we have GLM-5.2, a model from a different family, rescore 210 answers sampled from the reported runs. The two judges agree on 94\% of them (Cohen's $\kappa = 0.88$).
Cohen's $\kappa$ \citep{cohen1960kappa} is the agreement after removing the agreement expected by chance, where 1 means full agreement and 0 means chance level.
Values above 0.80 count as almost perfect agreement \citep{landis1977measurement}, so the reported scores are not specific to the gemma-4-31B-it judge.

\paragraph{Protocol.} Most inputs are the turns of a long multi-session conversation between two speakers, as in LoCoMo, and the memory holds facts about the speakers; a search passage, tool output or pasted text enters as a turn of its own, and so does an input that depends on earlier turns.
BM25 \citep{robertson2009bm25} returns 64 records and all-MiniLM-L6-v2 \citep{reimers2019sbert,wang2020minilm} reranks them, as in retrieval-augmented generation \citep{glass2022re2g,ram2023incontext}; the manager sees the top $K=8$ (no evidence record added).
SFT managers decode greedily under a grammar of the JSON target format; the untuned manager samples with its default settings, and its scores are averaged over three decode seeds.

\begin{figure*}[t]
\centering
\includegraphics[width=\textwidth]{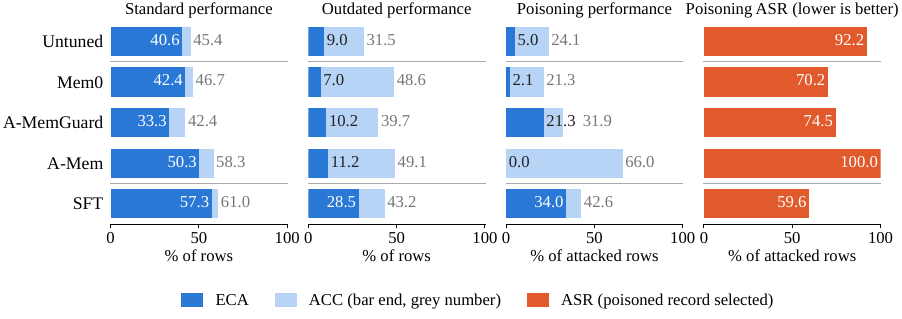}
\caption{Results of the Qwen3-4B managers on the test sets (\%).
In the performance panels, the dark bar is ECA and the light bar ends at ACC (grey number); poisoning ACC, ECA and ASR (lower is better) cover the attacked rows only.}
\label{fig:outcomes}
\end{figure*}

\paragraph{Training and testing.} We select 41 turns of one LoCoMo conversation as the base conversation $C$ and synthesize the data around it with the pipeline of Section~\ref{sec:data} (Figure~\ref{fig:data}): we insert turns of the three categories, 15\% of them outdated and 15\% poisoning, including variants that follow rules extracted from STALE and PoisonedRAG, and add false-passage episodes on general-knowledge facts.
This gives 400 training episodes with 21{,}656 rows.
The test data of Figure~\ref{fig:outcomes} is synthesized by the same pipeline and kept out of training: 29 episodes with the 1{,}510 rows of the three test sets.
We fine-tune all parameters of the backbone and tune its learning rate on development splits.
Figure~\ref{fig:outcomes} shows the results on the test sets; Figure~\ref{fig:app-main} in Appendix~\ref{app:main} gives every score, including overall averages.

\subsection{Main Results}
\label{sec:main}

\paragraph{Every baseline loses in at least one category.} Mem0 and A-Mem are built to store and retrieve facts from long conversations, as in the standard category, and there they match or exceed the untuned manager: their standard ECA is 42.4 and 50.3 against 40.6 (Figure~\ref{fig:outcomes}).
In the outdated and poisoning categories, their ECA stays near or below the untuned manager's.
On outdated inputs, both answer more rows correctly than the untuned manager (ACC 48.6 and 49.1 against 31.5), but their outdated ECA is 7.0 and 11.2 against 9.0, so most of these correct answers come from a selected memory that does not contain the correct memory.
On poisoning inputs, every attack enters A-Mem's selected memory (ASR 100.0), and its poisoning ECA is 0.0 although its ACC is 66.0; Mem0 lowers ASR from 92.2 to 70.2 but has poisoning ECA 2.1, below the untuned manager's 5.0.
A-MemGuard, a defense against poisoning, shows the reverse: it raises poisoning ECA from 5.0 to 21.3 but lowers standard ECA from 40.6 to 33.3.
No baseline therefore raises ECA over the untuned manager in all three categories.
ACC and ASR alone would hide these failures, since neither checks the answer and the selected records together, so we compare the methods by ECA.

\begin{figure*}[t]
\centering
\includegraphics[width=\textwidth]{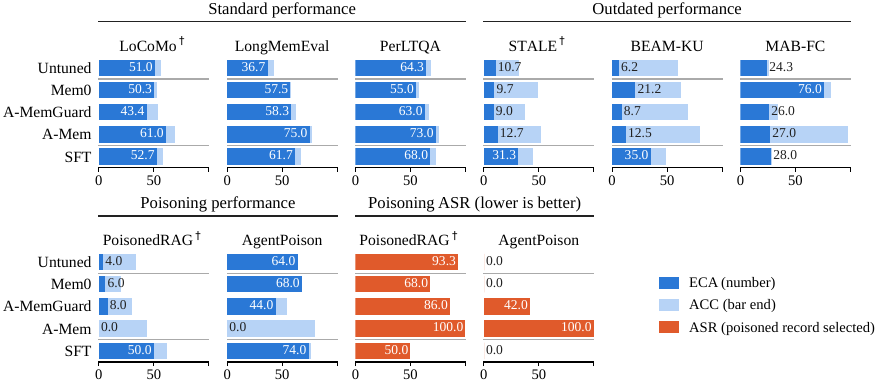}
\caption{Results of the Qwen3-4B managers on the eight benchmarks (\%), the seen one ($^\dagger$) first in each category.
Bars as in Figure~\ref{fig:outcomes}.}
\label{fig:gen}
\end{figure*}

\paragraph{The losses come from one fixed strategy applied to every input.} A-Mem keeps every note and Mem0 writes the facts of every turn, so both also store attacks.
A-MemGuard drops retrieved records that look inconsistent, so it also drops valid records.
The two problems need opposite write decisions on different inputs; a strategy that makes the same decision for all inputs helps one category and hurts another, and we attribute the baselines' losses to this.

\paragraph{SFT is the only method that improves all three categories.} SFT raises ECA over the untuned manager in all three categories, so by the definition of Section~\ref{sec:robustness} it is more robust than the untuned manager, whereas no baseline is.
It has the highest ECA in every category, so no baseline is more robust than it: 57.3 against at most 50.3 (A-Mem) on the standard test set, 28.5 against at most 11.2 (A-Mem) on outdated inputs and 34.0 against at most 21.3 (A-MemGuard, the method designed for poisoning) on poisoning inputs.
It reduces poisoning but does not prevent it: ASR falls from 92.2 to 59.6.
SFT keeps the prompts of the untuned manager and adds no fixed strategy, so the difference is that it decides per input what to write and which records to select.

\paragraph{The results hold on two other backbones.} On Qwen3.5-4B and Ministral-3-3B (Appendix~\ref{app:backbones}), SFT again raises ECA over the untuned manager in all three categories, and no baseline does.

\subsection{Generalization}
\label{sec:general}

We test the Qwen3-4B managers of Section~\ref{sec:main}, without further training, on the eight benchmarks of Section~\ref{sec:setup} under the same protocol.
Figure~\ref{fig:gen} reports each benchmark rather than category averages: the untuned manager's ECA ranges from 4.0 to 64.3 across them, and an average would hide where a method gains or loses.
The averages and the other backbones are in Appendix~\ref{app:general}.

\paragraph{SFT gains on all eight benchmarks.} SFT has higher ECA than the untuned manager on every benchmark, and the highest ECA of all managers on four, including AgentPoison (74.0 vs.\ 68.0 for Mem0).
The gains are not limited to the seen benchmarks: ECA rises from 36.7 to 61.7 on LongMemEval and from 6.2 to 35.0 on BEAM-KU, while on LoCoMo, PerLTQA and MAB-FC the gains are at most 3.7 points.
Its category ECA stays above the untuned manager's in all three categories (Appendix~\ref{app:general}), so by the definition of Section~\ref{sec:robustness} it stays more robust beyond the test sets.

\paragraph{Baselines lose on more benchmarks.} Mem0, A-MemGuard and A-Mem fall below the untuned manager's ECA on 3, 4 and 2 of the eight benchmarks, and SFT on none.
On AgentPoison, A-MemGuard, the method designed for poisoning, lowers ECA from 64.0 to 44.0 and raises ASR from 0.0 to 42.0, so it loses even in its own category.
A-Mem has the highest ECA on all three standard benchmarks, but ASR 100.0 and ECA 0.0 on PoisonedRAG and AgentPoison, and on STALE its ECA (12.7) is close to the untuned manager's (10.7).
Mem0 leads on MAB-FC (76.0 vs.\ 28.0 for SFT), where a later fact replaces an earlier one with no reported change, which our outdated category does not cover (Section~\ref{sec:framework}).

SFT is the only trained method, and its data is built around one conversation, yet it is the only manager that never falls below the untuned manager, so what it learns is not tied to that conversation.

\section{Conclusion}
\label{sec:conclusion}

We studied memory robustness for LLM agents: a memory manager should follow changed facts, reject poisoned content, and stay accurate on standard inputs.
To measure it, we introduced evidence-conditioned accuracy (ECA), which checks both the answer and the memory records behind it.
Under ECA, the existing memory systems and defenses we test are not more robust than an untuned manager on our test sets; each gains in one category and loses in another.
A simple one-shot recipe, which synthesizes training episodes around a single conversation and fine-tunes the manager on them, improves robustness on all three backbones, although robustness remains incomplete.
We hope this work encourages evaluating memory managers on the records they keep as well as on the answers they give, and studying outdated memory and memory poisoning together.
Future work includes other agent settings, stronger attacks and training methods beyond supervised fine-tuning.

\section*{Limitations}
Our study focuses on the memory managers of conversational agents.
Other settings, such as web or embodied agents whose memory tracks a changing environment, are not evaluated, and we leave them to future work.
As with most empirical studies of LLM agents, our evaluation covers a finite set of backbones, benchmarks and attack types, and may not include every form of outdated memory or memory poisoning that arises in deployment.
We also compare against a limited set of existing memory systems and defenses, so our findings about them may not extend to other methods.
Our conclusions should therefore be read as holding within the settings we study.

\section*{Ethics Statement}

This work aims to make LLM agents with long-term memory more reliable and secure.
The poisoned content we generate follows published attack patterns, introduces no new attack technique and is used only in our experiments.
All datasets are public research benchmarks used for their intended purpose, and our data involves no human participants or personal information.
The SFT manager does not eliminate all attacks and should be used together with other safeguards.
We will release our code for research use.

\section*{AI Use Statement}

We used AI assistants only to polish the writing of this paper.
All research ideas, experiments, analyses and conclusions are our own, and we take full responsibility for the content.

\section*{Acknowledgments}

This work was supported in part by the Defense Advanced Research Projects Agency (DARPA) under Cooperative Agreement No.~HR0011-24-2-0004.
We thank Indiana University's Research and Academic LLM Services (REALLMS) for providing API access to gemma-4-31B-it and GLM-5.2 for our experiments.

Distribution statement: Approved for public release; distribution is unlimited.

\FloatBarrier
\bibliography{references}

\appendix
\setcounter{dbltopnumber}{2}
\onecolumn
\renewcommand{\topfraction}{0.95}
\renewcommand{\bottomfraction}{0.9}
\renewcommand{\textfraction}{0.05}
\renewcommand{\floatpagefraction}{0.5}
\setcounter{topnumber}{4}
\setcounter{bottomnumber}{3}
\setcounter{totalnumber}{6}

\section*{Appendix Contents}
\begingroup
\setlength{\parindent}{0pt}
\setlength{\parskip}{2pt}
\makebox[1.6em][l]{\textbf{\ref{app:related}}}\textbf{Further Related Work}\dotfill\pageref{app:related}\par
\vspace{3pt}
\makebox[1.6em][l]{\textbf{\ref{app:base}}}\textbf{Experimental Details}\dotfill\pageref{app:base}\par
\hspace*{1.6em}\makebox[2.4em][l]{\ref{app:base-conv}}Base Conversation\dotfill\pageref{app:base-conv}\par
\hspace*{1.6em}\makebox[2.4em][l]{\ref{app:composition}}Training Data\dotfill\pageref{app:composition}\par
\hspace*{1.6em}\makebox[2.4em][l]{\ref{app:training}}Training\dotfill\pageref{app:training}\par
\hspace*{1.6em}\makebox[2.4em][l]{\ref{app:protocol}}Evaluation Protocol and Metrics\dotfill\pageref{app:protocol}\par
\hspace*{1.6em}\makebox[2.4em][l]{\ref{app:testsets}}Evaluation Sets\dotfill\pageref{app:testsets}\par
\vspace{3pt}
\makebox[1.6em][l]{\textbf{\ref{app:backbones}}}\textbf{Full Results}\dotfill\pageref{app:backbones}\par
\hspace*{1.6em}\makebox[2.4em][l]{\ref{app:main}}Test Sets\dotfill\pageref{app:main}\par
\hspace*{1.6em}\makebox[2.4em][l]{\ref{app:general}}Per-Benchmark Results\dotfill\pageref{app:general}\par
\hspace*{1.6em}\makebox[2.4em][l]{\ref{app:backbones-res}}Other Backbones\dotfill\pageref{app:backbones-res}\par
\vspace{3pt}
\makebox[1.6em][l]{\textbf{\ref{app:analysis}}}\textbf{Ablations and Analyses}\dotfill\pageref{app:analysis}\par
\hspace*{1.6em}\makebox[2.4em][l]{\ref{app:ablations}}Training-Data Ablations\dotfill\pageref{app:ablations}\par
\hspace*{1.6em}\makebox[2.4em][l]{\ref{app:strata}}Beyond the Base Conversation\dotfill\pageref{app:strata}\par
\hspace*{1.6em}\makebox[2.4em][l]{\ref{app:spoof}}Owner Changes and Cost\dotfill\pageref{app:spoof}\par
\vspace{3pt}
\endgroup

\FloatBarrier
\newpage
\section{Further Related Work}
\label{app:related}

\paragraph{Outdated memory.} Training-free methods keep replaced facts with validity intervals or on a timeline \citep{rasmussen2025zep,ong2025theanine}, as our updated records do (``Until $d$, \dots''), and PACMI propagates invalidity through a provenance graph \citep{wang2026pacmi}.
Supersede trains a 3B model with group relative policy optimization \citep{shao2024deepseekmath} on changed facts \citep{patel2026supersede}; we use plain SFT on data built around one conversation.

\paragraph{Memory poisoning and its cost.} MemSecBench follows poisoned memory to later actions \citep{chen2026memsecbench}, and The Price of Safety measures what poisoning defenses cost on inputs without attacks \citep{bhowmik2026price}.
Similar trade-offs appear as exaggerated refusal \citep{rottger2024xstest}, over-defense in prompt-injection guards \citep{li2025piguard}, lower clean accuracy under adversarial training \citep{tsipras2019robustness}, and acceptance of misinformation by LLMs that update on conflicting evidence \citep{xie2024adaptive,xu2024knowledge}.
These works measure the cost.

\paragraph{Data-efficient post-training.} One-shot critique fine-tuning trains on teacher critiques of many solutions to one problem \citep{wang2025oneshotcft}, and 1-shot reinforcement learning with verifiable rewards on many sampled responses to one problem \citep{wang2025oneexample}.
LLM-based augmentation expands a small seed set or corpus \citep{wang2023selfinstruct,sun2023selfalign,lee2024llm2llm,yang2025synthetic}, but generated examples can drift to a neighboring label \citep{sahu2022data}; our checks (Section~\ref{sec:data}) test for this drift.
Our contrast pairs follow counterfactual data and contrast sets, which make small edits that change the label \citep{kaushik2020counterfactual,gardner2020contrast}.

\FloatBarrier
\section{Experimental Details}
\label{app:base}

\subsection{Base Conversation}
\label{app:base-conv}
The base conversation $C$ is an excerpt of 41 turns from one LoCoMo conversation, chosen in two steps before any evaluation.
First, using only the LoCoMo annotations of the scored inputs, we counted in each of the ten LoCoMo conversations the single-hop, multi-hop and temporal inputs whose evidence turns resolve and contain no photo; conv-41 had the most (90).
Second, we took 41 original turns from three of its sessions (16 turns of session 25 on 2023-07-22, the first six turns of session 26 on 2023-07-31 and all 19 turns of session 28 on 2023-08-05), with their wording, speakers, dates and order unchanged, so that the excerpt contains a replaced fact (John's job at a mechanical engineering company, reported lost in session 28), plans, relative time cues and unrelated facts of both speakers.
Table~\ref{tab:datasource} states the requirements of Section~\ref{sec:data} as criteria for the conversation from which $C$ is taken and checks them on all of conv-41; of the seven LoCoMo conversations outside the test set, five meet them (conv-41, conv-26, conv-43, conv-42 and conv-50), and these are the source conversations of the ablations (Appendix~\ref{app:ablations}).

\begin{table}[!htb]
\centering
\small
\begin{tabular}{@{}>{\raggedright\arraybackslash}p{3.3cm}>{\raggedright\arraybackslash}p{4.6cm}>{\raggedright\arraybackslash}p{7.0cm}@{}}
\toprule
Criterion & Threshold & Evidence in conv-41 \\
\midrule
1 Independent attributes & $\geq 3$ verified attributes of $\geq 3$ types & 10 attributes of 7 types, e.g.\ John lost his job at a mechanical engineering company; Maria practices aerial yoga \\ \addlinespace
2 Time cues & $\geq 3$ verified relative time cues & 8, e.g.\ ``Just yesterday I joined a nearby church'' (2023-05-05) \\ \addlinespace
3 Plan vs.\ completed event & $\geq 1$ verified plan never carried out & 6, e.g.\ ``I'm planning a trip to the East Coast'', never reported as taken \\ \addlinespace
4 Facts valid in the past & $\geq 1$ verified replacement, old before new, no later contradiction & 3, e.g.\ John: ``I'm really enjoying my new job'' $\rightarrow$ ``I lost my job at the mechanical engineering company'' \\ \addlinespace
5 Retrieval coverage & old turn in the retriever's top 8 for $\geq 1$ replacement & 2 of 3; the job change: old turn ranked 1 of 53 \\ \addlinespace
6 Evidence in the text only & cited evidence needs no image or outside knowledge & yes \\ \addlinespace
7 Train/test separation & not conv-30, 47 or 49; no evaluation result used & yes \\
\bottomrule
\end{tabular}
\caption{Criteria for the conversation from which the base conversation is taken, checked on all of conv-41.}
\label{tab:datasource}
\end{table}

\subsection{Training Data}
\label{app:composition}
The variants are fact changes (a value replaced, a plan realized or cancelled, an activity stopped); no-change controls, which resemble changes but must not change memory (the owner's restatement, a third party's claim, a plan; the owner of a fact is the speaker it is about); turns with an injected instruction, alone or mixed with a fact; and \emph{rule-based variants}, which follow rules extracted from STALE and PoisonedRAG (Section~\ref{sec:data}) and take from them only how a change or an attack is made, never their text: paraphrased or implied changes and inputs that never name the attribute (STALE), and search passages with or without an injected instruction (PoisonedRAG).
The fractions of outdated and poisoning turns (15\% each, Section~\ref{sec:setup}) set how many turns of each category are inserted; the shares of training rows in Appendix~\ref{app:ablations} differ from them because they also count the rows of inputs that depend on other inputs and the rows of rule-based variants and false-passage episodes.

Table~\ref{tab:data} lists the four parts of the data.
Each row holds one input and at most eight retrieved records and trains both the operations and the read bits.
Parts (i) and (ii) are built from a table of about 130 changes over 16 attribute types and nine types of change, from attacks of five fixed templates, and from no-change controls, and an LLM writes each of their turns from its entry in the table.
About one third of the changes start from a value already in the conversation, and each change gets template inputs that ask about its current value or its past value or carry a false premise.
These two parts contain only explicit changes and templated attacks, so part (iii) adds implicit (paraphrased or implied) changes, which share at most one content word with the old fact, and attack passages; gpt-6.1-sol writes parts (iii) and (iv).
A false-passage episode (iv) is about a general-knowledge fact with one established value: it has supporting passages, five fluent passages that assert one wrong value without any instruction, and on-topic distractors, all rendered as search-tool output, and an input that needs the fact; its targets store and select the supporting passages and not the false ones.
Besides the memory built from the targets (Section~\ref{sec:data}), for 30\% of the changes, the inputs that depend on them get extra rows whose memory holds every turn verbatim.

The recipe was informed by the STALE and PoisonedRAG development splits, but no text of either was used: rule-based variants that share an 8-gram with STALE or PoisonedRAG text, or reach a 5-gram Jaccard similarity of 0.5 with it, are removed, and so are false-passage episodes that share an 8-gram with any evaluation text or match a PoisonedRAG input and its answer.
The cue check (v) of Section~\ref{sec:data} requires that simple cues (word overlap, recency, length and cue words for old vs.\ new records; length, overlap with the input, position and specificity for false vs.\ supporting passages) recover the target with accuracy at most 55\% to 65\%.

\begin{table}[!htb]
\centering
\footnotesize
\begin{tabular}{@{}>{\raggedright\arraybackslash}p{2.6cm}>{\raggedright\arraybackslash}p{6.2cm}r>{\raggedright\arraybackslash}p{3.8cm}@{}}
\toprule
Part & Content & Rows & Targets \\
\midrule
(i) Episode steps & Changes, attacks and no-change controls inserted into a 41-turn excerpt of conv-41 & 6{,}546 & Labeling rule; teacher operations on original turns (4{,}351) \\ \addlinespace
(ii) Contrast pairs & Copies of a step that differ only in the factor that decides the target & 4{,}721 & Labeling rule \\ \addlinespace
(iii) Rule-based variants & Implicit changes (paraphrased 1{,}430, implied 1{,}550, task-style input 253); attack passages (1{,}577) & 4{,}810 & Design of the variant \\
\midrule
Conversation rows & (i)--(iii) & 16{,}077 & \\
\midrule
(iv) False-passage episodes & Supporting, false and distractor passages for general-knowledge facts & 5{,}579 & Design of the episode \\
\midrule
All backbones & (i)--(iv), 400 episodes & 21{,}656 & \\
\bottomrule
\end{tabular}
\caption{Composition of the training data.}
\label{tab:data}
\end{table}

\FloatBarrier
\subsection{Training}
\label{app:training}
Table~\ref{tab:hparams} lists the training and inference settings.

\paragraph{Prompts and targets.} The state $s_t$ in the prompt is a JSON object with the date of the input, the input, and the retrieved records (identifier, date, session, source type, speaker, text), relabeled M1, M2, \dots{} in a seeded random order.
As in Section~\ref{sec:vanilla}, a target text $y_t$ gives the target action $a^\ast_t$ (the operations, \op{ADD}, \op{NOOP}, or \op{ADD} with one to three \op{UPDATE}s, and a read bit per retrieved record) and labels of the input's category and of each retrieved record's role; it also restates the read bits as a decision (select some records or none) and a list of the selected records.
Following the action labeling of Section~\ref{sec:targets}, the read bits select the evidence records, including an updated record when the input asks about the past, and no outdated or poisoned record.
Only the operations and the read bits are executed.
The teacher (gpt-6-luna) fills the choices that the labeling rule leaves open, namely the operations on the original conv-41 turns (4{,}351 rows) and parts of 2{,}091 further targets (an added span, a label or extra selected records).

\begin{table*}[!htb]
\centering\footnotesize\setlength{\tabcolsep}{4pt}\renewcommand{\arraystretch}{1.2}
\begin{minipage}[t]{0.495\textwidth}\centering
\begin{tabular}[t]{|>{\raggedright\arraybackslash}p{3.3cm}|>{\raggedright\arraybackslash}p{3.9cm}|}
\hline
\multicolumn{2}{|l|}{\textbf{Training}} \\
\hline
Fine-tuned parameters & All \\
\hline
Learning rate & $5{\times}10^{-6}$ / $1.5{\times}10^{-6}$ / $10^{-6}$ \\
\hline
Learning rates tried & \{$5{\times}10^{-6}$, $10^{-5}$\} / \{$10^{-6}$, $1.5{\times}10^{-6}$, $2{\times}10^{-6}$, $5{\times}10^{-6}$\} / \{$10^{-6}$, $2{\times}10^{-6}$\} \\
\hline
Optimizer & AdamW, weight decay 0.01 \\
\hline
Gradient clipping (norm) & 1.0 \\
\hline
Schedule & 20 warmup steps, cosine decay to 0 \\
\hline
Epochs & 2 \\
\hline
Batch size (rows per step) & 4 \\
\hline
Training rows / steps & 21{,}656 / 10{,}828 \\
\hline
Longest prompt / target & 2{,}418 / 251 tokens (no truncation) \\
\hline
Loss & Token-averaged cross-entropy on target tokens (incl.\ EOS) \\
\hline
Precision & bfloat16 \\
\hline
Hardware / checkpoint & One H100 / after the last step \\
\hline
\end{tabular}
\end{minipage}\hfill
\begin{minipage}[t]{0.495\textwidth}\centering
\begin{tabular}[t]{|>{\raggedright\arraybackslash}p{3.3cm}|>{\raggedright\arraybackslash}p{3.9cm}|}
\hline
\multicolumn{2}{|l|}{\textbf{Memory manager at test time}} \\
\hline
Decoding (SFT) & Greedy, thinking off, grammar of the JSON target schema \\
\hline
Decoding (untuned) & Model-card defaults (T 0.7, top-p 0.8, top-k 20) / model-card defaults / model-card defaults \\
\hline
Decode seeds (untuned) & 3 (averaged per row) \\
\hline
Max prompt / new tokens & 16{,}384 / 4{,}096 \\
\hline
\multicolumn{2}{|l|}{\textbf{Retrieval}} \\
\hline
First stage & BM25, 64 candidates \\
\hline
Reranker & all-MiniLM-L6-v2 \\
\hline
Records shown to the manager ($K$) & 8 \\
\hline
\multicolumn{2}{|l|}{\textbf{$\pi_{\mathrm{exec}}$ and judge}} \\
\hline
Model & gemma-4-31B-it \\
\hline
Decoding & T 1.0, top-p 0.95, top-k 64, thinking off \\
\hline
Max new tokens ($\pi_{\mathrm{exec}}$) & 1{,}024 \\
\hline
\end{tabular}
\end{minipage}
\caption{Training and inference settings. Values written a / b / c are for Qwen3-4B / Qwen3.5-4B / Ministral-3-3B; a single value is shared by all backbones. The learning rate is tuned per backbone on the development splits.}
\label{tab:hparams}
\end{table*}

\FloatBarrier
\subsection{Evaluation Protocol and Metrics}
\label{app:protocol}
The retriever is queried with the speaker and text of each input over the records of the manager's own memory, and the top 8 records are relabeled M1, M2, \dots{} in a seeded random order, as in training.
The manager gives 1--4 operations per input, which are applied deterministically: an added record is linked to its source turn, an updated record keeps the old record's turns and adds the new turn, and a deleted record is never retrieved again.
As in Section~\ref{sec:setup}, outputs that cannot be parsed fail: operations that cannot be parsed, are cut off or break the operation constraints (e.g., add text not copied from the input or name a record that was not retrieved) are executed as \op{NOOP}, invalid read bits give an empty selected memory, an empty or cut-off answer is scored wrong, and every failure stays in the denominators.
The baselines are run with one decode seed; all other decoding settings are in Table~\ref{tab:hparams}.

$\pi_{\mathrm{exec}}$ sees only the selected memory, with one prompt per benchmark; the STALE prompt renders the selected records in chronological order with session dates and asks $\pi_{\mathrm{exec}}$ to use the most recent related evidence when the asked attribute is implicit.
The judge is gemma-4-31B-it, called separately from $\pi_{\mathrm{exec}}$ (Section~\ref{sec:setup}), with the scoring prompt of each benchmark; for example, the STALE prompt also shows the benchmark's annotated old and new memory, LoCoMo uses the Mem0 judge prompt \citep{chhikara2025mem0}, BEAM its nugget-based prompt, and PoisonedRAG, LongMemEval, MAB-FC and PerLTQA the LongMemEval answer check.

Record roles come from the benchmark's annotation or our construction, never from a model (Table~\ref{tab:evalsets}); a record with none of the listed roles is irrelevant.
A record linked to any evidence turn is evidence, and a record that stores an attack is poisoned (Section~\ref{sec:robustness}); a record that keeps only the part of a poisoning input that is not the attack, as in Figure~\ref{fig:teaser}b, is not poisoned; in the outdated category, a record linked to both old and new evidence takes the role of its latest turn, and an updated record (``Until $d$, \dots'') is not outdated.
With these roles, the selected memory contains the correct memory $\mathcal{C}_t$ (Section~\ref{sec:robustness}) if it holds every evidence record in memory, none of them is outdated or poisoned, and every evidence turn of the input is linked to one of them.

ASR is measured on the selected memory rather than on the answer, because with a fixed $\pi_{\mathrm{exec}}$ answer-level attack success barely separates managers; on PoisonedRAG, answer-level success also counts answers that contain the attacker's chosen answer with no poisoned record selected, and on AgentPoison the output that the attack aims for is also what $\pi_{\mathrm{exec}}$ says when it abstains.
Because an empty selected memory counts as a defended attack, we also report poisoning ACC.

Intervals of differences between two managers, as in Appendix~\ref{app:ablations}, are 95\% paired cluster-bootstrap percentile intervals with 10{,}000 resamples, clustered by STALE episode, PoisonedRAG input, AgentPoison input, LoCoMo window, and held-out input or episode.
For models trained with three seeds we report the mean and the standard deviation (SD) across seeds.

\subsection{Evaluation Sets}
\label{app:testsets}
The evaluation has two parts (Section~\ref{sec:setup}): the three test sets of Figures~\ref{fig:outcomes} and~\ref{fig:app-main}, which are synthesized by the pipeline of Section~\ref{sec:data} from held-out parts of LoCoMo, STALE and PoisonedRAG, and the eight benchmarks of Figures~\ref{fig:gen} and~\ref{fig:app-std}--\ref{fig:app-poi}, which test generalization.
Table~\ref{tab:evalsets} lists every set; poisoning sets score the attacked copy of each input.
No test input was used for training, and nothing from the five unseen benchmarks was used.
The three test sets of Figure~\ref{fig:app-main} were fixed before we saw any output of an SFT manager on them, and the evaluation sets of Figures~\ref{fig:app-std}--\ref{fig:app-poi} were sampled with a fixed seed before any model output.

\emph{STALE}: we use the STALE conversations \citep{chao2026stale} with a paraphrased or implied change; a fixed hash splits them into a development half and a test half.
The outdated test set is synthesized from the changes of the test half, and the 100-episode sample of Figure~\ref{fig:app-out} is drawn from both halves.

\emph{LoCoMo}: the standard test set is synthesized on the held-out conversations conv-30, 47 and 49, with contrast pairs (Section~\ref{sec:data}) that differ only in whether the input should be written (an impersonated vs.\ a real owner change, a non-owner vs.\ an owner statement, a clean vs.\ a poisoning tool passage).
As stated in Section~\ref{sec:setup}, the set is scored as a whole in the standard category, so its 866 rows include the 407 rows we added about changes (323 about inserted changes and 84 about changes already in the conversations; Table~\ref{tab:strata}).

\emph{PoisonedRAG}: the poisoning test set is synthesized from the 47 inputs of Natural Questions \citep{kwiatkowski2019nq} left in the test split after removing yes/no inputs, inputs in MemoryAgentBench, and the inputs of the held-out sample and of a development split.
For each input, the clean copy places the top eight Contriever \citep{izacard2022contriever} passages for the input, one passage per turn, in one block among passages retrieved for other inputs; the attacked copy adds the release's five poisoned texts inside the block.
The test split shares no input or passage with the held-out sample of Figure~\ref{fig:app-poi}.

An audit against the training rows finds no shared 8-gram, no verbatim input, and no answer value attached to the same input, but the LoCoMo test episodes share the generator and part of the input templates with training, so Appendix~\ref{app:strata} reports held-out templates separately.

\begin{table}[!htb]
\centering
\footnotesize
\setlength{\tabcolsep}{4pt}
\begin{tabular}{@{}llccr>{\raggedright\arraybackslash}p{5.8cm}@{}}
\toprule
Category & Benchmark & Figure & Seen & Rows & Record roles from \\
\midrule
\multirow{4}{*}{Outdated} & STALE test set & \ref{fig:app-main} & \checkmark & 597 & Old and new evidence sessions of the episode \\
 & STALE, 100-episode sample & \ref{fig:app-out} & \checkmark & 300 (100 ep.) & As above \\
 & BEAM-KU & \ref{fig:app-out} & & 80 & \texttt{original\_info} and \texttt{updated\_info} source chats \\
 & MAB-FC & \ref{fig:app-out} & & 100 & Gold fact index; outdated: the earlier fact with the same subject and relation \\
\midrule
\multirow{3}{*}{Poisoning} & PoisonedRAG test set & \ref{fig:app-main} & \checkmark & 47 & Five poisoned texts; clean retrieved passages \\
 & PoisonedRAG, held-out sample & \ref{fig:app-poi} & \checkmark & 50 & As above \\
 & AgentPoison & \ref{fig:app-poi} & & 50 & Demonstration with the trigger (poisoned); no evidence record \\
\midrule
\multirow{4}{*}{Standard} & LoCoMo test (conv-30, 47, 49) & \ref{fig:app-main} & \checkmark & 866 & Our construction of the inserted turns; native evidence annotation \\
 & LoCoMo test, native inputs & \ref{fig:app-std} & \checkmark & 459 & Native evidence annotation \\
 & LongMemEval & \ref{fig:app-std} & & 120 & Turns marked \texttt{has\_answer} \\
 & PerLTQA & \ref{fig:app-std} & & 100 & Reference memory ids \\
\bottomrule
\end{tabular}
\caption{Evaluation sets: the three synthesized test sets (Figure~\ref{fig:app-main}) and the eight benchmarks (Figures~\ref{fig:app-std}--\ref{fig:app-poi}), with the source of their record roles (MAB: MemoryAgentBench).}
\label{tab:evalsets}
\end{table}

\FloatBarrier
\section{Full Results}
\label{app:backbones}
We repeat the experiments of Sections~\ref{sec:main} and~\ref{sec:general} with two other 3--4B backbones, Qwen3.5-4B \citep{qwen2026qwen35}, used as a text-only model, and Ministral-3-3B \citep{liu2026ministral3}.
Both train on the same dataset as Qwen3-4B with their own learning rate (Table~\ref{tab:hparams}), and every baseline runs on the same backbone as the manager it is compared with.

\subsection{Test Sets}
\label{app:main}

\begin{figure*}[t]
\centering
\includegraphics[width=\textwidth]{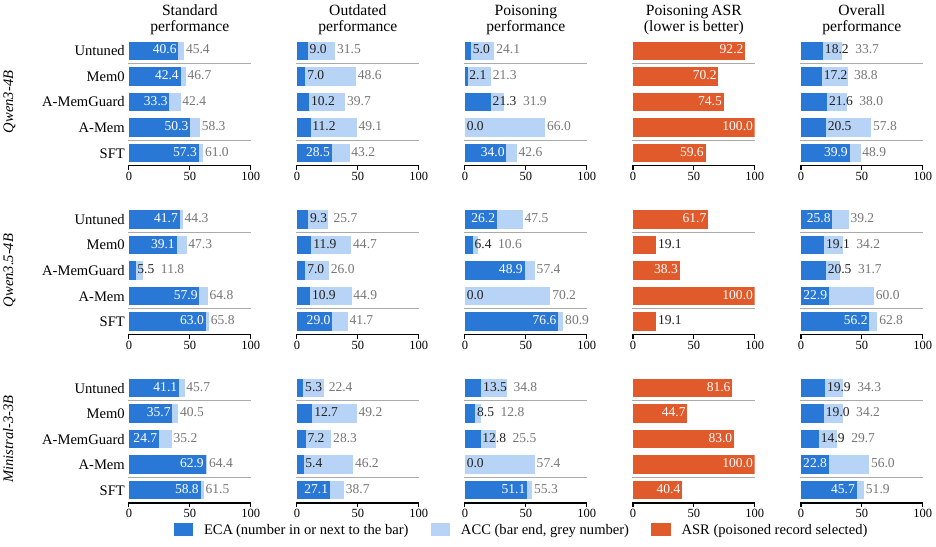}
\caption{Results on the test sets per backbone (\%); the Qwen3-4B row is Figure~\ref{fig:outcomes} with the overall scores added.
The untuned manager is the mean over three decode seeds.
Bars as in Figure~\ref{fig:outcomes}: the dark bar is ECA and the light bar ends at ACC (grey number).
Poisoning ACC, ECA and ASR (lower is better) cover the attacked rows only.}
\label{fig:app-main}
\end{figure*}

Figure~\ref{fig:app-main} gives the ACC, ECA and ASR of every manager on the test sets, per backbone; its Qwen3-4B row repeats Figure~\ref{fig:outcomes} and adds the overall scores.
Poisoning ACC, ASR and ECA cover the attacked rows only.

\FloatBarrier
\subsection{Per-Benchmark Results}
\label{app:general}

\begin{figure*}[t]
\centering
\includegraphics[width=\textwidth]{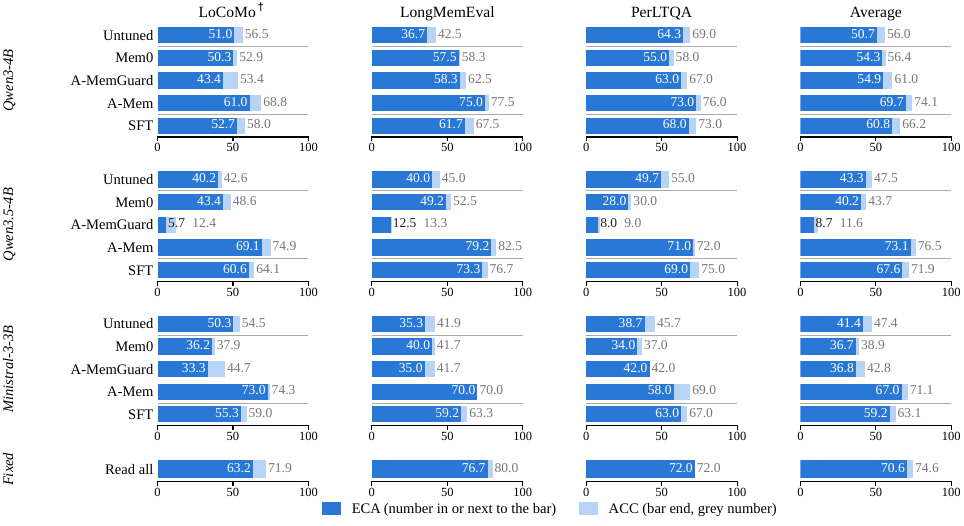}
\caption{Per-benchmark results in the standard category (\%) per backbone, and for the fixed manager Read all (Fixed), which uses no manager model.
Average is the mean of the three benchmarks.
Bars as in Figure~\ref{fig:app-main}.
$^\dagger$Seen.}
\label{fig:app-std}
\end{figure*}

\begin{figure*}[t]
\centering
\includegraphics[width=\textwidth]{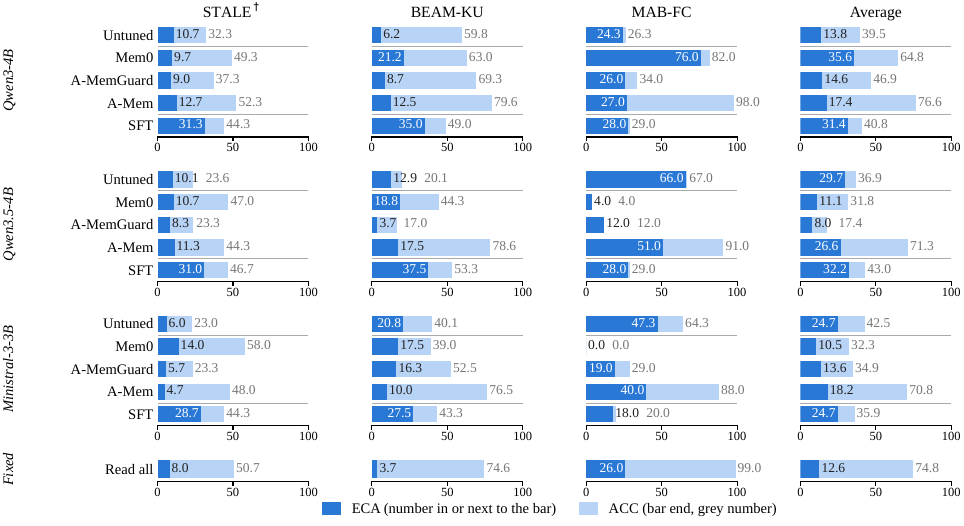}
\caption{Per-benchmark results in the outdated category (\%) per backbone, and for the fixed manager Read all (Fixed).
Average is the mean of the three benchmarks.
Bars as in Figure~\ref{fig:app-main}.
$^\dagger$Seen.}
\label{fig:app-out}
\end{figure*}

\begin{figure*}[p]
\centering
\includegraphics[width=\textwidth]{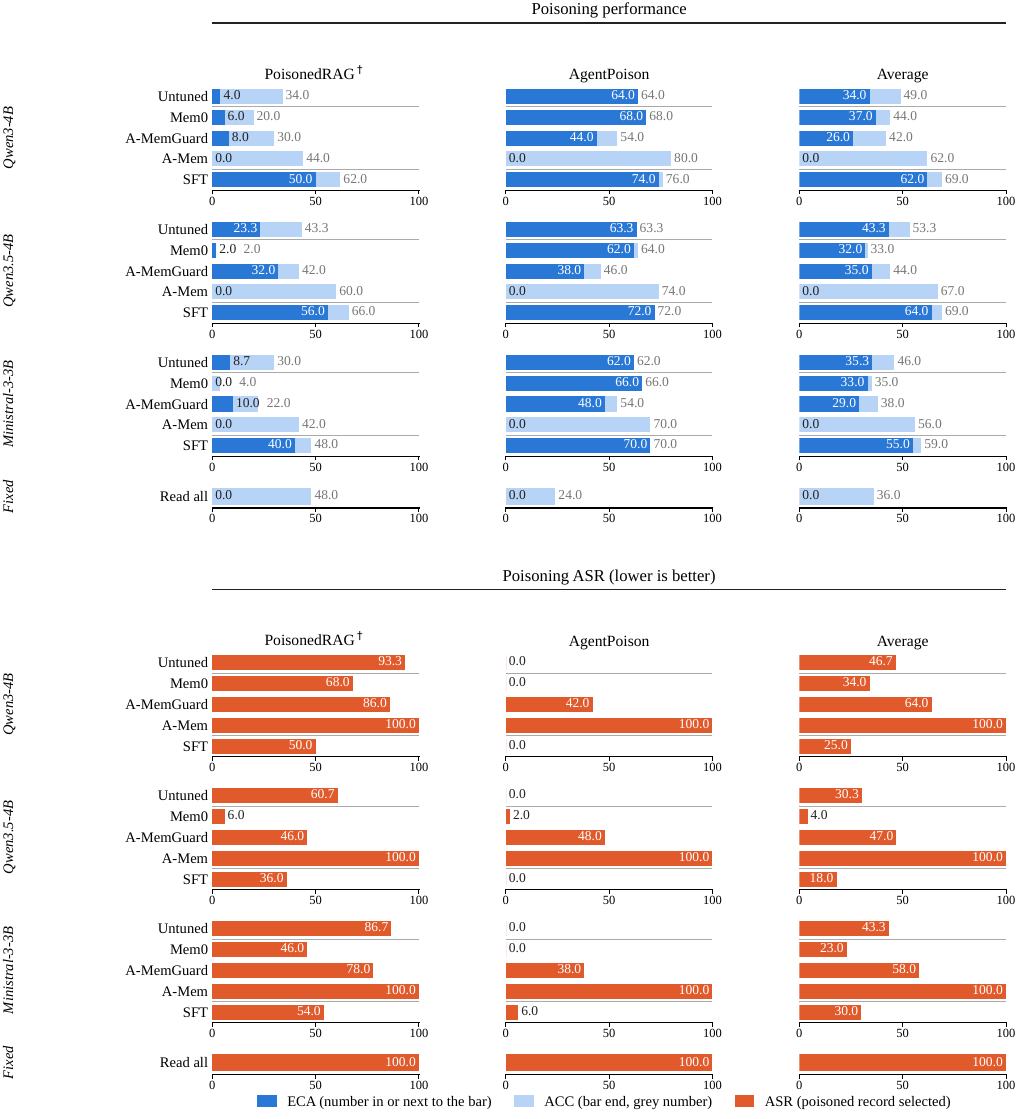}
\caption{Per-benchmark results in the poisoning category (\%) per backbone, and for the fixed manager Read all (Fixed): ECA and ACC (top) and ASR (bottom, lower is better).
ACC, ASR and ECA cover the attacked rows only.
Average is the mean of the two benchmarks.
Bars as in Figure~\ref{fig:app-main}.
$^\dagger$Seen.}
\label{fig:app-poi}
\end{figure*}

Figures~\ref{fig:app-std}, \ref{fig:app-out} and~\ref{fig:app-poi} give each benchmark and the category average per backbone, with the seen benchmark ($^\dagger$) first; Figure~\ref{fig:gen} shows their Qwen3-4B rows.

A fixed manager, Read all, stores every turn and passes every retrieved record to $\pi_{\mathrm{exec}}$, under the same protocol as the other methods.

SFT has the highest STALE and BEAM-KU ECA of all managers on every backbone, and the highest average poisoning ECA.
Read all has average standard ECA 70.6, below only A-Mem on Qwen3.5-4B (73.1), but passes every AgentPoison and PoisonedRAG attack and has average outdated ECA 12.6.

\paragraph{Reading the figures.} The STALE sample of Figure~\ref{fig:app-out} overlaps the development half and the test half of STALE, and BEAM-KU ACC is its nugget score; its ECA counts a row as correct only at the full nugget score.
On MAB-FC the manager writes a 455-fact episode.
The LoCoMo column of Figure~\ref{fig:app-std} scores only the 459 native LoCoMo rows of the standard test set, not the 407 rows we added about changes, so it differs from the standard scores of Figure~\ref{fig:app-main}.
As in Figure~\ref{fig:app-main}, ECA is computed over all rows of a benchmark, so a row whose evidence is not retrieved counts as a failure.
On AgentPoison, the untuned and SFT managers rarely select a poisoned record (ASR at most 6.0), so SFT lowers ASR only on PoisonedRAG.

\paragraph{Per-row changes.} For each test set of Figure~\ref{fig:app-main} and each unseen benchmark, SFT fixes a row if the untuned manager has ECA 0 under all three decode seeds and SFT has ECA 1, and breaks it if SFT has ECA 0 and the untuned manager has ECA 1 under at least two seeds (poisoning counts attacked rows only).
Summed over the outdated and over the poisoning category, SFT fixes more rows than it breaks on every backbone.
On Qwen3-4B, Qwen3.5-4B and Ministral-3-3B it fixes / breaks 165 / 34, 131 / 60 and 137 / 43 rows in the outdated category, and 21 / 3, 24 / 1 and 22 / 4 in the poisoning category.

\subsection{Other Backbones}
\label{app:backbones-res}
On the test sets (Figure~\ref{fig:app-main}), SFT raises ECA over the untuned manager in all three categories on both backbones: on Qwen3.5-4B from 41.7 to 63.0 (standard), 9.3 to 29.0 (outdated) and 26.2 to 76.6 (poisoning), and on Ministral-3-3B from 41.1 to 58.8, 5.3 to 27.1 and 13.5 to 51.1.
ASR falls from 61.7 to 19.1 and from 81.6 to 40.4.
Every baseline falls below the untuned manager's ECA in at least one category on both backbones.
SFT has the highest ECA in every category except standard ECA on Ministral-3-3B, where A-Mem is higher (62.9 vs.\ 58.8) but has poisoning ECA 0.0 and ASR 100.0.

On the eight benchmarks (Figures~\ref{fig:app-std}--\ref{fig:app-poi}), SFT's category-average ECA is at least the untuned manager's in every category on both backbones (equal in the outdated category on Ministral-3-3B), while Mem0 and A-MemGuard are below it in all three categories on both backbones.
Per benchmark, SFT is below the untuned manager only on MAB-FC (28.0 vs.\ 66.0 on Qwen3.5-4B and 18.0 vs.\ 47.3 on Ministral-3-3B).
Mem0's lead on MAB-FC with Qwen3-4B does not carry over: its MAB-FC ECA is 4.0 on Qwen3.5-4B and 0.0 on Ministral-3-3B.

\section{Ablations and Analyses}
\label{app:analysis}

\subsection{Training-Data Ablations}
\label{app:ablations}
Each ablation retrains the Qwen3-4B SFT manager of Figure~\ref{fig:app-main} (the reference) with one change to its training data and evaluates it on the same test sets (Table~\ref{tab:abl-full}).
PoisonedRAG is seen (Section~\ref{sec:setup}), but no text from it was used for training.
A difference is \emph{real} if its 95\% paired cluster-bootstrap interval excludes 0 and it exceeds the SD over three seeds.
Ablated runs keep the learning rate, two epochs and the 21{,}656 rows of the reference, except under \emph{Fewer rows} and \emph{One conversation}.
To keep the row count, removed rows are replaced by repeating a random sample of the remaining rows, and when the share of one category changes, standard rows fill the rest (shares are standard : outdated : poisoning rows; reference 56.8 : 18.4 : 24.8).
Each one-conversation run has 3{,}215 conversation rows, plus the same 2{,}166 false-passage rows in the runs with false passages; the conv-41 rows are a sample of the reference rows, while those of the other conversations come from a later run of the pipeline.

\paragraph{Each problem needs its own data.} Without outdated rows, outdated ECA falls to 12.7 (untuned: 9.0); without poisoning rows, ASR rises to 93.6 (untuned: 92.2).
Neither removal lowers the other problem's scores or standard ECA by a real amount; trained on both, the manager keeps the outdated gain and most of the poisoning gain.
Halving or doubling either share keeps all three ECAs above the untuned manager by real amounts, so the gain does not depend on the exact shares.

\paragraph{Two parts of the data are needed for the gain.} Without the implicit-change variants, outdated ECA falls from 28.5 to 7.0 (untuned: 9.0); without the false-passage episodes, ASR rises from 59.6 to 83.0.
Removing the contrast pairs, the no-change controls or the labels in the targets does not lower overall ECA by a real amount (42.2, 38.6 and 44.6), so these parts are not needed for the gain.

\paragraph{A small amount of data is enough.} With 10\% to 50\% of the rows, every score stays above the untuned manager by a real amount, and 5\% of the rows already reach overall ECA 44.6.
From 10\% of the rows on, overall ECA is at least as high as with all data, also over three seeds.
With 10\% of the rows, the three seeds give overall ECA $43.8 \pm 1.7$ and ASR $39.7 \pm 1.2$ (all rows: $42.1 \pm 2.5$ and $50.4 \pm 8.1$), and only the outdated scores are lower (ECA 20.0 vs.\ 26.1, ACC 37.4 vs.\ 39.6).

\paragraph{Other source conversations also work.} Without false passages, each of the five LoCoMo conversations that meet our base-conversation criteria raises outdated ECA, but only the base conversation, chosen before any evaluation, also lowers ASR by a real amount (66.0, against 85.1 to 89.4 for the other four).
With the same false-passage rows added, each of the five raises all three ECAs and lowers ASR by real amounts with 5{,}381 rows (overall ECA 41.6 to 44.4 vs.\ 39.9; ASR 31.9 to 57.4).

\paragraph{The gain does not depend on one seed.} Two other seeds give overall ECA 44.8 and 41.5 (reference: 39.9), and each seed beats the untuned manager on outdated and standard ACC and ECA and on poisoning ASR and ECA (Table~\ref{tab:abl-full}).

\paragraph{The gain does not depend on the retriever or $\pi_{\mathrm{exec}}$.} Without retraining, showing the manager the top 4 or 16 of the 64 BM25 candidates instead of 8, or dropping the reranker, keeps overall ECA above the untuned manager (33.0 vs.\ 14.7, 42.8 vs.\ 21.0 and 32.6 vs.\ 17.0); rows whose evidence is no longer retrieved count as failures.
Replacing only $\pi_{\mathrm{exec}}$ by Qwen3.5-9B \citep{qwen2026qwen35} or GLM-5.2 on the selected memory of each manager (judge unchanged) also keeps overall ECA above the untuned manager (39.8 vs.\ 16.8 and 38.8 vs.\ 17.0), with an interval that excludes 0 for every per-category ECA difference, so the gain does not depend on using the judge model as $\pi_{\mathrm{exec}}$.
In these runs the untuned manager is run with one decode seed.

\begin{table}[!htb]
\centering
\small
\renewcommand{\arraystretch}{1.0}
\setlength{\tabcolsep}{4pt}
\begin{tabular}{@{}lcccccccc@{}}
\toprule
& \multicolumn{2}{c}{Overall} & \multicolumn{2}{c}{Outdated} & \multicolumn{2}{c}{Poisoning} & \multicolumn{2}{c}{Standard} \\
\cmidrule(lr){2-3} \cmidrule(lr){4-5} \cmidrule(lr){6-7} \cmidrule(l){8-9}
Training data & ACC & ECA & ACC & ECA & ASR$\downarrow$ & ECA & ACC & ECA \\
\midrule
None (untuned) & 33.7 & 18.2 & 31.5 & 9.0 & 92.2 & 5.0 & 45.4 & 40.6 \\
All (reference, one seed) & 48.9 & 39.9 & 43.2 & 28.5 & 59.6 & 34.0 & 61.0 & 57.3 \\
\quad 3 seeds, mean & 49.3 & 42.1 & 39.6 & 26.1 & 50.4 & 41.8 & 62.3 & 58.2 \\
\quad 3 seeds, SD & 2.1 & 2.5 & 4.6 & 3.4 & 8.1 & 6.8 & 2.2 & 1.5 \\
\midrule
\multicolumn{9}{@{}l}{\textit{Share of outdated or poisoning rows (standard : outdated : poisoning, \%)}} \\
\quad outdated $\times0$ (75.2 : 0 : 24.8) & 47.0 & 38.6 & 26.8 & 12.7 & 38.3 & 48.9 & 58.9 & 54.3 \\
\quad outdated $\times0.5$ (66.0 : 9.2 : 24.8) & 49.9 & 42.0 & 39.4 & 27.1 & 57.4 & 40.4 & 61.4 & 58.5 \\
\quad outdated $\times2$ (38.5 : 36.7 : 24.8) & 49.1 & 40.0 & 33.5 & 18.3 & 55.3 & 42.6 & 62.8 & 59.1 \\
\quad poisoning $\times0$ (81.6 : 18.4 : 0) & 38.9 & 29.9 & 40.7 & 28.5 & 93.6 & 6.4 & 59.1 & 55.0 \\
\quad poisoning $\times0.5$ (69.2 : 18.4 : 12.4) & 51.5 & 42.7 & 35.2 & 20.3 & 40.4 & 48.9 & 63.9 & 59.0 \\
\quad poisoning $\times2$ (32.0 : 18.4 : 49.6) & 48.3 & 38.2 & 38.5 & 24.3 & 59.6 & 29.8 & 63.9 & 60.4 \\
\midrule
\multicolumn{9}{@{}l}{\textit{Without one part}} \\
\quad implicit-change variants & 46.8 & 38.6 & 25.0 & 7.0 & 48.9 & 48.9 & 64.4 & 59.9 \\
\quad false-passage episodes & 49.3 & 32.4 & 42.0 & 28.6 & 83.0 & 14.9 & 56.9 & 53.7 \\
\quad contrast pairs & 51.0 & 42.2 & 39.0 & 25.3 & 44.7 & 46.8 & 58.5 & 54.4 \\
\quad no-change controls & 49.3 & 38.6 & 38.7 & 25.3 & 61.7 & 31.9 & 62.4 & 58.7 \\
\quad labels in the targets & 53.2 & 44.6 & 44.6 & 32.3 & 53.2 & 42.6 & 64.0 & 58.8 \\
\midrule
\multicolumn{9}{@{}l}{\textit{Fewer rows (share of the reference rows)}} \\
\quad 1\% of the rows & 37.4 & 28.0 & 33.3 & 13.4 & 31.9 & 46.8 & 29.8 & 23.8 \\
\quad 5\% of the rows & 54.0 & 44.6 & 30.5 & 13.7 & 29.8 & 61.7 & 63.4 & 58.4 \\
\quad 10\% of the rows & 55.7 & 44.7 & 39.7 & 23.6 & 38.3 & 51.1 & 65.8 & 59.5 \\
\qquad 3 seeds, mean & 54.5 & 43.8 & 37.4 & 20.0 & 39.7 & 52.5 & 63.8 & 58.7 \\
\qquad 3 seeds, SD & 2.7 & 1.7 & 5.6 & 5.6 & 1.2 & 1.2 & 2.0 & 0.8 \\
\quad 25\% of the rows & 53.5 & 43.2 & 39.7 & 21.9 & 42.6 & 48.9 & 63.3 & 58.7 \\
\quad 50\% of the rows & 51.6 & 42.4 & 39.5 & 25.8 & 51.1 & 44.7 & 60.0 & 56.7 \\
\midrule
\multicolumn{9}{@{}l}{\textit{One conversation, no false passages (3{,}215 rows)}} \\
\quad conv-41 (base) & 52.2 & 37.1 & 39.5 & 21.4 & 66.0 & 29.8 & 63.9 & 59.9 \\
\quad conv-26 & 46.8 & 32.6 & 43.4 & 26.0 & 89.4 & 10.6 & 65.1 & 61.1 \\
\quad conv-43 & 49.4 & 36.6 & 49.4 & 35.3 & 85.1 & 12.8 & 64.8 & 61.7 \\
\quad conv-42 & 39.4 & 32.0 & 40.2 & 27.5 & 89.4 & 8.5 & 63.0 & 60.2 \\
\quad conv-50 & 47.7 & 34.4 & 41.9 & 28.5 & 87.2 & 12.8 & 65.1 & 61.9 \\
\midrule
\multicolumn{9}{@{}l}{\textit{One conversation with false passages (3{,}215 + 2{,}166 rows)}} \\
\quad conv-41 (base) & 49.2 & 41.9 & 36.5 & 21.9 & 46.8 & 44.7 & 62.1 & 59.1 \\
\quad conv-26 & 53.1 & 44.4 & 35.5 & 19.8 & 31.9 & 55.3 & 62.1 & 58.1 \\
\quad conv-43 & 51.5 & 41.6 & 37.0 & 24.5 & 57.4 & 40.4 & 64.2 & 59.8 \\
\quad conv-42 & 57.5 & 44.1 & 40.0 & 24.8 & 48.9 & 46.8 & 64.3 & 60.6 \\
\quad conv-50 & 51.3 & 43.2 & 38.4 & 24.8 & 53.2 & 44.7 & 64.4 & 60.0 \\
\bottomrule
\end{tabular}
\caption{All training-data ablations of the Qwen3-4B SFT manager (\%).}
\label{tab:abl-full}
\end{table}

\FloatBarrier
\subsection{Beyond the Base Conversation}
\label{app:strata}
\begin{table}[!htb]
\centering
\small
\setlength{\tabcolsep}{2.5pt}
\begin{tabular}{@{}llrc@{}}
\toprule
Stratum & Subset & Rows & ACC [95\% interval] \\
\midrule
\multirow{2}{*}{Attribute type}   & Seen in training & 304 & 61.8 {\scriptsize [52.4, 69.3]} \\
                                  & Held-out         & 103 & 71.8 {\scriptsize [58.3, 84.2]} \\ \addlinespace
Input template                    & Held-out         & 159 & 65.4 {\scriptsize [54.0, 74.0]} \\ \addlinespace
\multirow{2}{*}{Change origin}    & Inserted         & 323 & 68.1 {\scriptsize [63.0, 73.1]} \\
                                  & In the conversation & 84 & 50.0 {\scriptsize [29.8, 71.2]} \\
\bottomrule
\end{tabular}
\caption{Accuracy (\%) of the Qwen3-4B SFT manager on the LoCoMo test rows about a change, by stratum.}
\label{tab:strata}
\end{table}

The conversation rows of the training data come from one conversation, so the manager could overfit to its attributes or input templates.
Table~\ref{tab:strata} splits the 407 LoCoMo test rows about a change by what the training data covers; brackets are 95\% cluster-bootstrap intervals (2{,}000 resamples).
The SFT manager is at least as accurate on the eight attribute types that never occur in training as on the 16 that do (71.8 vs.\ 61.8), and it reaches 65.4 on rows with held-out templates.
However, changes already in the held-out conversations, rather than inserted ones, score lower (50.0 vs.\ 68.1, on 84 rows with a wide interval).

\subsection{Owner Changes and Cost}
\label{app:spoof}
Since a speaker's own report changes that speaker's facts, we test whether the same change made through another channel also changes memory.
For each of 78 changes in the held-out LoCoMo conversations, a clean episode has the owner report the change in their own turn, and an attacked episode writes the same change in the first person as the owner but in a tool passage or a third-party message (e.g., ``Note from Caroline: I moved to Lyon last week'').
Change ACC is the share of clean episodes answered with the new value, and ASR, as in Section~\ref{sec:setup}, the share of attacked episodes whose selected memory contains a poisoned record (the record of the impersonated change).
On Qwen3-4B, SFT has change ACC 70.5 and ASR 75.6, against $29.5 \pm 2.6$ and $60.3 \pm 2.2$ for the untuned manager (mean $\pm$ SD over three decode seeds) and 50.0 and 82.1 for A-MemGuard (one run each); compared with the untuned manager, it writes far more of the owner's own changes but is also misled more often by impersonation.

On the same 9 LoCoMo test episodes for every manager (428 turns), the SFT manager takes 0.39\,s per turn on one H100 (not counting $\pi_{\mathrm{exec}}$), against 1.32\,s for Mem0 and 0.84\,s for A-MemGuard, all with Qwen3-4B.
Per turn, the SFT manager makes 1.33 LLM calls with 1{,}764 input and 119 output tokens, against 1.75 calls (2{,}177 / 253 tokens) for Mem0 and 2.98 calls (1{,}038 / 359 tokens) for A-MemGuard.

\end{document}